\documentclass[
aps,
prfluids,
amsmath,amssymb,superscriptaddress
]{revtex4}
\usepackage{rotating, graphicx}
\usepackage{color}
\graphicspath{{./graph/}}
\usepackage{color}
\usepackage{caption,subcaption}
\usepackage[font=Large]{subcaption}
\pdfoutput=1
\begin{document}
\title{Numerical simulations of a falling film on the inner surface of a rotating cylinder}
\author{U.~Farooq}
\email{usmaan.farooq13@imperial.ac.uk}
\affiliation{Department of Chemical Engineering, Imperial College London, SW7 2AZ, United Kingdom}
\author{J.~Stafford}
\affiliation{Department of Chemical Engineering, Imperial College London, SW7 2AZ, United Kingdom}
\affiliation{School of Engineering, University of Birmingham, B15 2TT, United Kingdom}
\author{C.~Petit}
\affiliation{Department of Chemical Engineering, Imperial College London, SW7 2AZ, United Kingdom}
\author{O.~K.~Matar}
\email{o.matar@imperial.ac.uk}
\affiliation{Department of Chemical Engineering, Imperial College London, SW7 2AZ, United Kingdom}

\begin{abstract}
A flow in which a thin film falls due to gravity on the inner surface of a vertical, rotating cylinder is investigated. This is performed using two-dimensional (2D) and three-dimensional (3D) direct numerical simulations, with a volume-of-fluid approach to treat the interface. The problem is parameterised by the Reynolds, Froude, Weber, and Ekman numbers. The variation of the Ekman number $(Ek)$, defined to be proportional to the rotational speed of the cylinder, has a strong effect on the flow characteristics. %
Simulations are conducted over a wide range of $Ek$ values ($0\leq Ek \leq 484$) in order to provide %
detailed insight into how this parameter influences the flow. Our results indicate that increasing $Ek$, which leads to a rise in the magnitude of centrifugal forces, 
produces a stabilising effect, suppressing wave formation. Key flow features, such as the transition from a 2D to a more complex 3D wave regime, %
are %
influenced significantly by this stabilisation, and are investigated in detail. Furthermore, the imposed rotation results in distinct flow characteristics such as the development of angled waves, which arise due to the combination of gravitationally- and centrifugally-driven motion in the axial and azimuthal directions, respectively. We also use a weighted residuals integral boundary layer method to determine a boundary in the space of Reynolds and Ekman numbers that represents a threshold beyond which waves have recirculation regions. %
\\
\end{abstract}
\maketitle

\section{Introduction}
Falling liquid films are of central importance to a range of industrial applications and associated unit operations, which include reactors, distillation columns, heat exchangers, condensers, and evaporators. 
It is unsurprising, therefore, that they have enjoyed significant attention in the literature for several decades since the seminal work of Kapitza %
\cite{kapitza_1_1948,kapitza_2_1949,Craster_Matar_2009,kalliadasis_springer_2012}.
Falling film flows are charactersised by complex dynamics and pattern formation. Waves emerge naturally from disturbances near the inlet, which then grow downstream and as their amplitude increases, nonlinearities then give rise to growth saturation; the resulting wave deceleration then leads to constant wave speed. Two distinct types of waves have been detected corresponding to short waves that are nearly sinusoidal in shape, typically found near the flow inlet, and longer, large-amplitude, solitary waves, further downstream; the latter have tall, well-separated peaks, which are preceded by capillary waves of much smaller amplitude and whose speed is maintained by the constant compression from the main solitary wave peak, located immediately upstream \cite{dietze_jfm_2016}.
The interfacial dynamics are also accompanied by transitions from two- to three-dimensional waves that resemble horseshoe-like coherent structures,   %
and at higher film Reynolds numbers, so-called `roll waves' overtake the capillary waves resulting in complex, 
apparently random, wave structures
\cite{patnaik_ijhff_1996,kalliadasis_springer_2012}.

Numerous methods have been employed to investigate the behaviour of thin falling films. Experimentally, film thicknesses and velocity profiles can be determined through laser-based techniques including fluorescence and particle image velocimetry measurements \cite{alekseenko_pf_2009, alekseenko_jfm_2012, liu_pf_1993, liu_jfm_1993, zadrazil_ijfm_2014a, zadrazil_ijfm_2014b, charogiannis_etfs_2015}. Furthermore, modelling and numerical simulations have been used extensively to provide insight into the complex falling film dynamics.  %
Low-dimensional (LD) and weighted integral boundary layer (WIBL) modelling have been used to provide an accurate representation of the hydrodynamics \cite{scheid_jfm_2006,kalliadasis_springer_2012}. Numerical modelling and direct numerical simulations (DNS) have also been deployed starting with the work of Ramaswamy {\it et al.} \cite{ramaswamy_jfm_1996} who were among the first to perform DNS on falling liquid films, using a finite-element method with a Lagrangian-Eulerian formulation to analyse the spatial and temporal stability of the flow.
Two-dimensional simulations of falling films using the volume-of-fluid method were performed by Gao {\it et al.} \cite{gao_jcp_2003} who examined the time-space wave evolution at different Reynolds and Weber numbers. Gao {\it et al.} \cite{gao_jcp_2003} and Nosoko and Miyara \cite{nosoko_pf_2004} also assessed the impact of forcing the inlet flow rate with certain frequencies on the emergent wave formation, an approach similar to that employed experimentally by Park {\it et al.} \cite{park_ijhmt_2004}. %
Recently, Denner {\it et al.} \cite{denner_jfm_2018} compared experimental measurements with DNS and LD modelling for solitary waves on inertia-dominated falling liquid films, finding good agreement. These authors have further investigated the onset of recirculation within the 
waves whose presence acts to intensify the rates of heat and mass transfer in the falling film \cite{roberts_ces_2000,dietze_jfm_2008, malamataris_pf_2008,albert_ijhmt_2014}.

In the present work, we consider the dynamics of a film falling under gravity on the inside of a cylinder, which is undergoing steady rotation; this study is carried out in connection with applications such as evaporators in which the rotation provides an additional degree of freedom to intensify heat and mass transfer rates \cite{stephan_htcb_1992}. 
Notably, it has been shown that the rotation increases the 
heat transfer coefficients by {25\%} in the case of a centrifugal thin film evaporator, whilst even greater increases are found in other geometries such as spinning disks \cite{yanniotis_ichmt_1996, chen_th_1997}. Furthermore, rotation provides greater control %
over the flow dynamics in comparison to the non-rotating falling film case. 
It is also noteworthy that the problem of a rotating, thin falling film is related to that involving %
a thin film flow down an inclined plane; in both flows, the films are influenced by a body force, which corresponds to centrifugation and gravity in the rotating and non-rotating cases, respectively.
In the inclined plane case, for angle $\gamma$ above the horizontal, the Kapitza instability, which eventually leads to wave formation, as described above, is present for  $Re > 5/6 \ \cot{\gamma}$ \cite{benjamin_jfm_1957, yih_pf_1963}. For angles past the vertical, the films, which are on the underside of an inclined plane, are also vulnerable to a Rayleigh-Taylor instability \cite{kondic_pof_2010}, as summarised in Figure 1 of Rietz {\it et al.} \cite{rietz_jfm_2017}. 

If one considers the forces acting on the system, the gravity component can be separated into a contribution in the streamwise direction and one normal to this into the plane. Inertia and the streamwise gravity component serve to destabilise the flow, whereas surface tension and the gravity component normal to the flow have a stabilising effect. This is comparable to the current case, in which the normal component of gravity in the inclined plane case plays the role of the centrifugal force due to rotation of the cylinder. Thus, the angle of inclination can be equated to the ratio of the centrifugal and gravitational accelerations.  %
Rietz {\it et al.} \cite{rietz_jfm_2017} have made use of this parameter to classify the results of their %
experimental study of thin film flow on the outside of a vertical, rotating cylinder that feature the formation of 2D and 3D waves, rivulet formation, and dripping. %

Linear stability analyses of a thin film on the surface of a rotating cylinder have been performed by Chen {\it et al.} \cite{chen_ijhmt_2004} and Davalos {\it et al.} \cite{davalos_pf_1993} using the lubrication approximation. Davalos {\it et al.} \cite{davalos_pf_1993} have noted that for flow on the inside of the cylinder, inertial, and capillary forces have a destabilizing effect, whereas the centrifugal force stabilizes the flow. From this analysis, a critical, so-called centrifugal number %
can be determined, suggesting that the flow is stable for a sufficiently large rotational speed. %

Although films falling on the exterior of a rotating cylinder have rich dynamics due to the simultaneous presence of Kapitza, Rayleigh-Taylor, and centrifugal instabilities \cite{taylor_prsl_1950, sharp_physica_1984,rietz_jfm_2017}, we focus on the effect of the stabilising centrifugal force associated with flow on the inner surface of a rotating cylinder; this force will act in competition with the destabilising gravitational force that leads to the Kapitza instability. 
Here, we will perform a
numerical investigation of the flow, %
which is yet to be studied in the nonlinear regime beyond the onset of linear instabilities %
\cite{davalos_pf_1993, ruiz_chavarria_jdp_1996, iwasaki_bjsme_1981}. %
The role that rotation has on the dynamics and stability of the flow will be examined and %
our results will demonstrate the emergence of large-amplitude waves that travel at a well-defined angle to the axis of the vertical cylinder. 

The rest of this paper is organized as follows: in Section \ref{section_2}, the problem formulation is presented, highlighting the key non-dimensional parameters associated with the flow via scaling of the governing equations; a brief exposition of the numerical methods deployed is also provided.
In Section \ref{section_3}, the results from two-dimensional simulations, within a rotating frame of reference, are presented, while in Section \ref{section_4}, the predictions from the three-dimensional simulations are discussed. %
Finally, conclusions and an outlook for future work are presented in Section \ref{section_5}.

\begin{figure}[ht]
	\centering
	\includegraphics[width=0.5\linewidth]{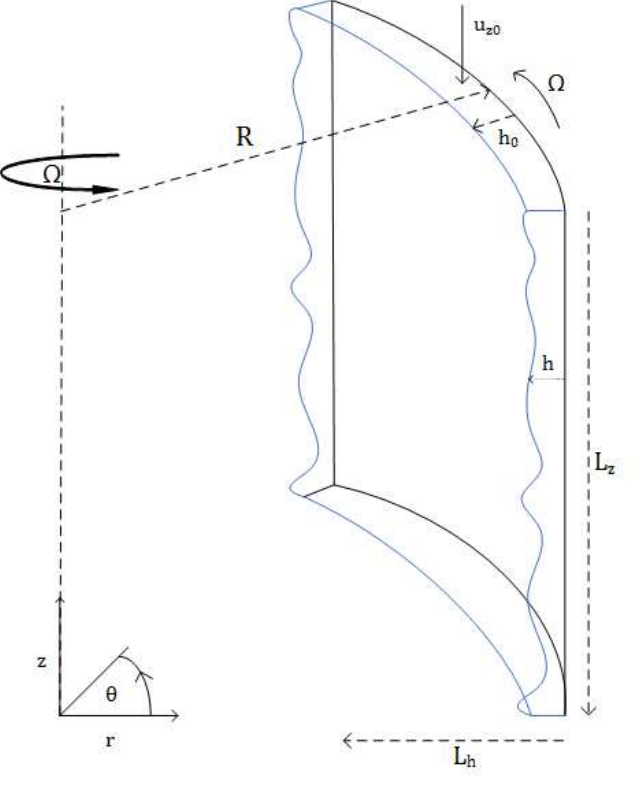}
	\caption{Schematic of the flow showing fluid entering through an inlet at the top of a cylinder of radius $R$, with velocity $u_{z0}$ and height $h_0$, adhering to the cylinder wall which rotates at constant speed $\Omega$, resulting in a film with height, $h = R - r$. The computational domain has lengths $L_z$ and $L_h$ in the axial and radial directions, and covers a 90$^\circ$ cylinder sector in the azimuthal direction. }
	\label{fig:Fig_1}
\end{figure}

\section{Problem formulation}\label{section_2}
\subsection{Governing equations}\label{g_eqn}
We consider a Newtonian liquid film of density $\rho_l$ and viscosity $\mu_l$ flowing due to gravity down the inner surface of a rigid and impermeable cylinder of radius $R$, oriented vertically, and rotating with a constant rotational speed $\Omega$. A gas phase of density $\rho_g$ and viscosity $\mu_g$ is also present in the cylinder, and the gas and liquid phases are separated by an interface with surface tension $\sigma$. We use  cylindrical co-ordinates, $r$, $\theta$, and $z$,  defined as shown in Fig \ref{fig:Fig_1}, with associated velocity components, $u_r$, $u_\theta$, and $u_z$, to describe the flow. The film has thickness $h(\theta,z,t)$ such that the gas-liquid interface is located at $r=R-h$. The fluid is considered to be incompressible, isothermal, and can be described by the continuity and momentum equations, respectively given by:
\begin{equation} \label{NS2_normal}
\nabla \cdot \mathbf{u}=0,
\end{equation}
\begin{equation} \label{NS1_normal}
\rho (\frac{\partial \mathbf{u}}{\partial t}+\mathbf{u} \cdot \nabla \mathbf{u})=-\nabla p +\mu \nabla^{2} \mathbf{u} + \rho \mathbf{g} + \sigma{\kappa}\mathbf{n}{\delta},
\end{equation}
where $\mathbf{u}$ is the velocity, $p$ the pressure, $\mathbf{g}$ the gravitational acceleration, $\kappa$, the curvature of the interface, $\mathbf{n}$ is the unit normal at the interface, and $\delta$ represents the Dirac delta function concentrated at the interface.

We use a volume-of-fluid (VOF) interface-capturing approach to simulate the interfacial dynamics within the open-source environment OpenFOAM. According to the VOF method, a species transport equation is used to determine the volume fraction, $\alpha$, of each phase in every computational cell. %
The function $\alpha$ allows one to define the local density and viscosity as 
\begin{equation} \label {rho_alpha}
\rho = \alpha\rho_g + (1-\alpha)\rho_l,
\end{equation}
\begin{equation} \label{mu_alpha}
\mu = \alpha \mu_g + (1-\alpha)\mu_l,    
\end{equation}
and $\alpha$ is advected using the following equation:
\begin{equation} \label {advection_eqn}
\frac{\partial \alpha}{\partial t}+\nabla \cdot(\alpha \boldsymbol{u})=0.
\end{equation}
We also use the continuum surface force approach to model the surface tension force according to which %
the normal $\mathbf{n}$ and curvature $\kappa$ are respectively expressed by
\begin{equation}
\mathbf{n}=\frac{\nabla\alpha}{|\nabla\alpha|},~~~ {\rm and} ~~~
\kappa=-\nabla\cdot \left(\frac{\nabla\alpha}{|\nabla\alpha|}\right).
\end{equation}

In order to construct a sharper interface, %
Eq. (\ref{advection_eqn}) is modified to compress the surface and reduce smearing; further details can be found in \cite{deshpande_csd_2012}. Roenby {\it et al.} \cite{roenby_rsp_2016}, recently implemented an isoAdvector scheme within OpenFOAM, allowing for higher Courant numbers than the standard solver with MULES. This uses the concept of isosurfaces to calculate more accurate face fluxes, specifically for the cells containing the interface. This geometric method has an optimum performance at $Co \approx 0.5$, compared to $Co \leq 0.1$ for the algebraic VOF approach implemented in the interFoam solver  \cite{roenby_rsp_2016}. This isoAdvector solver (interFlow) was used in the current study. 

The governing equations are rendered dimensionless via introduction of the following scaling
\begin{equation}\label{scalings}
\tilde{\textbf{u}}=\frac{\textbf{u}}{u_N}, \quad \tilde{\mathbf{x}}=\frac{\mathbf{x}}{h_N}, \quad \tilde{t}=\frac{t}{h_N/u_N}, \quad \tilde{p}=\frac{p}{\rho u_N^2},  \quad \tilde{\kappa}=\frac{\kappa}{1/h_N}, \quad %
\end{equation}
in which $h_N$ and $u_N$ correspond respectively to the Nusselt  thickness and velocity for a planar falling film in the absence of rotation: %
\begin{equation}
h_{N}=\sqrt[3]{\frac{3 \mu_{l} q_{N}}{\rho_{l} g}}, ~~~
u_{N}=\frac{\rho_{l} g h_{N}^{2}}{3 \mu_{l}};
\end{equation}
note that we have also set $\tilde{\delta}=h_N\delta$. 
In the above, %
the tildes designate the dimensionless variables. Applying these scalings to the mass and momentum equations, we obtain
\begin{equation}
\tilde{\nabla}\cdot {\tilde{\mathbf u}}=0,
\end{equation}
\begin{equation} \label{NS2_dim}
\frac{\partial \tilde{\mathbf{u}}}{\partial \tilde{t}}+\tilde{\mathbf{u}} \cdot  \tilde{\nabla} \tilde{\mathbf{u}}=-\tilde{\nabla} \tilde{p} + \frac{1}{Re}\tilde{\nabla}^{2} \tilde{\mathbf{u}} + \frac{1}{Fr^2} + \frac{1}{We}\tilde{\kappa}\tilde{\boldsymbol{\delta}},
\end{equation}
where the dimensionless parameters that appear in Eq. (\ref{NS2_dim}) correspond to the Reynolds, Froude, and Weber numbers, respectively given by:
\begin{equation}
Re = \frac{\rho u_Nh_N}{\mu}, \quad Fr = \frac{u_N}{\sqrt{gh_N}}, \quad We = \frac{\rho u_N^2 h_N}{\sigma}.
\end{equation}

At the cylinder surface, located at $\tilde{r}=1/\beta$, we impose a no-slip boundary condition such that the dimensionless azimuthal velocity is as follows:
\begin{equation}
\tilde{u}_{\theta} | _{r=R} = \frac{\Omega R}{u_N} = \frac{Ek}{Re},
\end{equation}
in which $Ek$ is the Ekman number given by
\begin{equation}
Ek = \frac{\rho (\Omega R) h_N}{\mu},
\end{equation}
and $\beta\equiv h_N/R$. %
We also impose continuous conditions at the gas boundary and outlet, and a steady, uniform velocity at the inlet. In dimensional terms, the flow is initiated with a film of thickness $h_N$ and velocity $u_N$ corresponding to the $Re$ given below. Finally, periodic boundary conditions are imposed in the azimuthal direction.

Equation (\ref{NS1_normal}) can be re-expressed as follows in a rotating reference frame:
\begin{equation}\label{rrfeqn}
\rho \left(\frac{\partial \mathbf{u}'}{\partial t}+\mathbf{u}' \cdot \nabla \mathbf{u}'\right)=-\nabla p +\mu \nabla^{2} \mathbf{u}' + \rho \mathbf{g} + \sigma{\kappa}\mathbf{n}{\delta} -\mathbf{\Omega} \times(\mathbf{\Omega} \times r)-2 \mathbf{\Omega} \times \mathbf{u}',
\end{equation}
where $\mathbf{\Omega} = -\Omega \mathbf{\hat{z}}$ and $\mathbf{\hat{z}}$ is the unit vector in the $z$ direction. Velocity is in the rotating reference frame such that $\mathbf{u}^{\prime}=\mathbf{u}-\Omega \mathbf{r}$. The fifth and sixth terms on the right-hand-side of Eq. (\ref{rrfeqn}) correspond to the centrifugal and Coriolis forces, respectively. 
Substitution of the 
scalings in Eq. (\ref{scalings}) into Eq. (\ref{rrfeqn}) yields
\begin{equation}
\label{rrfeqn_dless}
\frac{\partial \tilde{\mathbf{u}}'}{\partial \tilde{t}}+\tilde{\mathbf{u}}' \cdot \tilde{\nabla} \tilde{\mathbf{u}}'=-\tilde{\nabla} \tilde{p} + \frac{1}{Re}\tilde{\nabla}^{2} \tilde{\mathbf{u}}' + \frac{1}{Fr^2} + \frac{1}{We}\tilde{\kappa}\tilde{\boldsymbol{\delta}} 
-\frac{Ek}{Re}\left[\beta^2\left(\frac{Ek}{Re}\right)
(\mathbf{\hat{z}} \times (\mathbf{\hat{z}} \times \mathbf{\tilde{r}})) +2 (\mathbf{\hat{z}}  \times \mathbf{\tilde{u}}')
\right].
\end{equation}
A dimensionless domain size of $\tilde{L}_h \approx 7.5$ and $\tilde{L}_z \approx 725$ minimised the effect of the gas dynamics on the interface and ensured that there was sufficient space for the transition between wave regimes to occur unhindered.   %
Mesh refinement in the region of the film, $\frac{1}{\beta}-2.5 < \tilde{r} < \frac{1}{\beta}$ was performed such that the mesh size was 0.09 in the film region %
and 0.45 in $\tilde{z}$, %
broadly similar to the 2D domain employed by Gao {\it et al.} \cite{gao_jcp_2003}. 
The 3D case was constructed by extruding the 2D geometry and mesh in the azimuthal direction, producing a cylindrical sector with a mesh size of 0.46 in $\theta$ in the region of the film. A 90$^\circ$  sector of the cylinder was used with periodic boundary conditions so as to reduce the computational requirement. The choice of sector size was validated against numerical solutions obtained for a full cylinder. %
A dynamic time-step was selected such that the Courant number $Co < 0.5$ was satisfied as per the optimum performance of the solver \cite{roenby_rsp_2016}.

An air-water system was used such that the gas phase had density $\rho_g = 1.27 \ \text{kg.m}^{-3}$ and kinematic viscosity $\nu_g = 1.42 \times 10^{-5} \ \text{m}^2.\text{s}^{-1}$. The liquid film has density $\rho_l = 1000 \ \text{kg.m}^{-3}$, kinematic viscosity $\nu_l = 1.14 \times 10^{-6} \ \text{m}^2.\text{s}^{-1}$, and surface tension $\sigma_l = 7.28 \times 10^{-2} \ \text{N.m}^{-1}$. The dimensionless parameters based on these conditions are $Re = 53$, $Fr = 4.2$, $We = 0.18$, and %
$\beta=1.4\times 10^{-4}$ whilst $Ek$ varies between $0 \leq Ek \leq 484$. %

\section{Two-Dimensional Simulation} \label{section_3}
\begin{figure}
	\begin{minipage}{\textwidth}
		\begin{subfigure}{\linewidth}
			\centering
			\caption{\hspace*{-2em}}
			\includegraphics[width=0.6\textwidth]{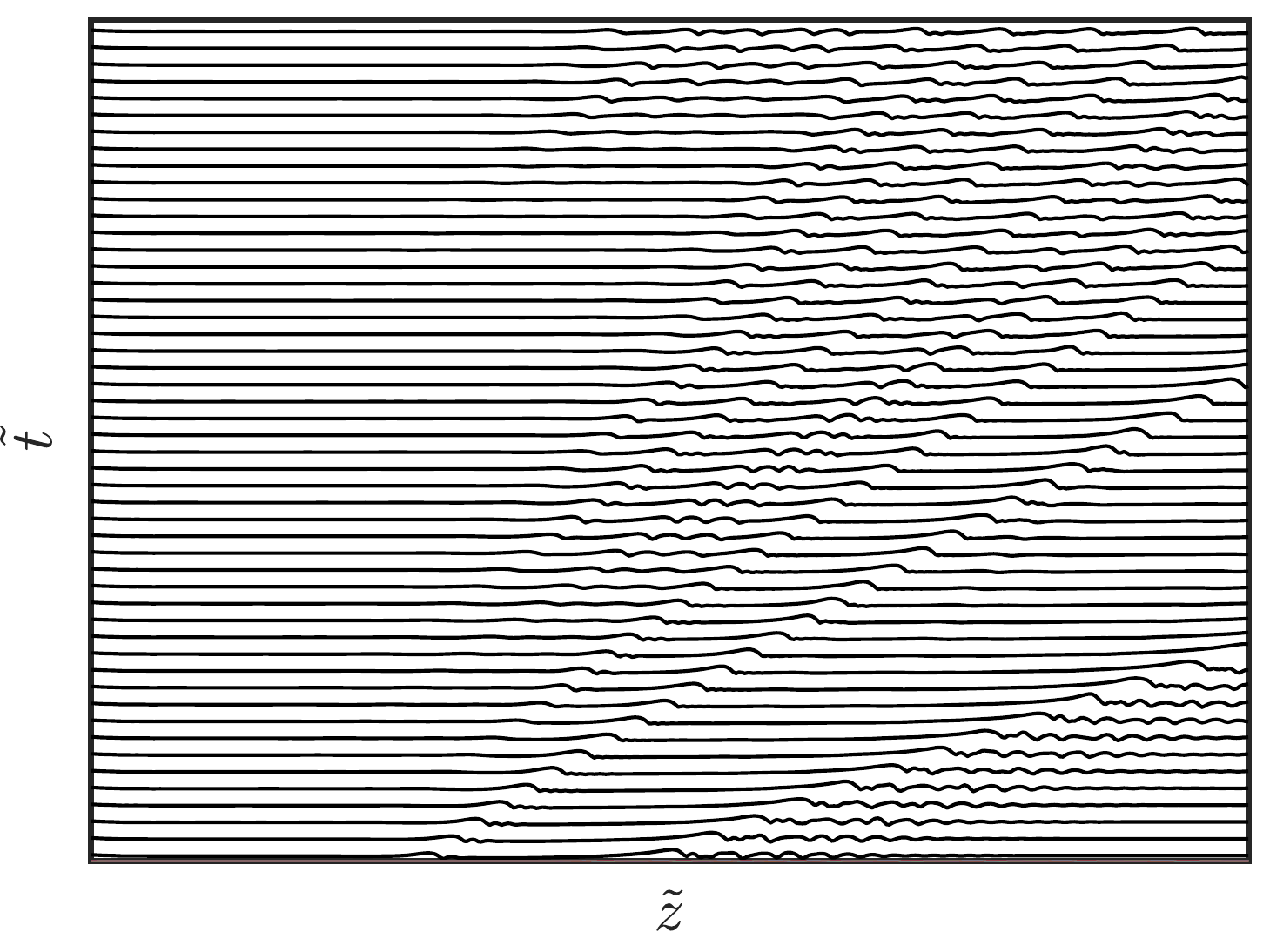}
			\label{fig:fig_2a}
		\end{subfigure}
	\end{minipage}
	\begin{minipage}{0.5\textwidth}
		\begin{subfigure}{\linewidth}
			\centering
			\caption{\hspace*{-1em} \vspace{0.5em}}
			\hspace{0.5cm}
			\vspace{2em}
			\includegraphics[width = 0.7\textwidth]{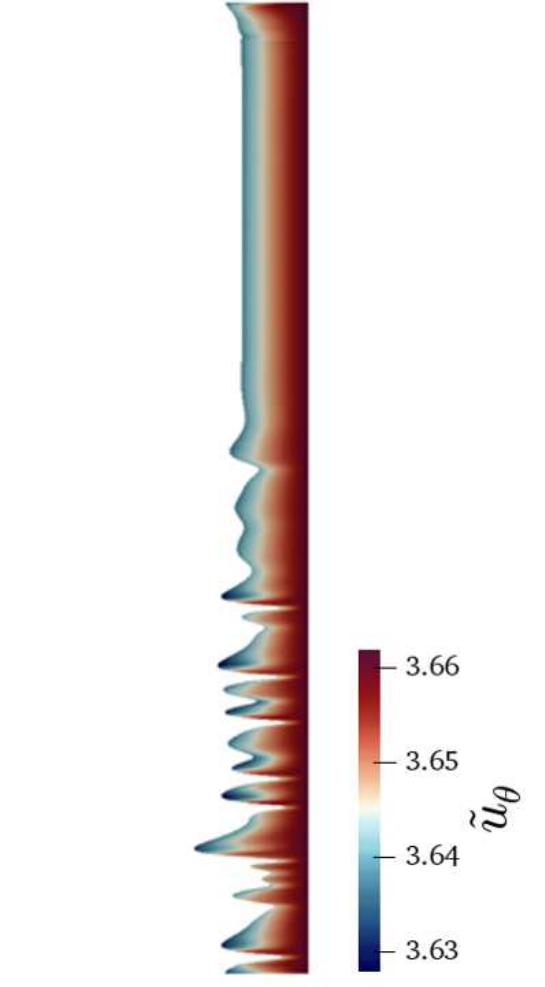}
			\label{fig:fig_2b}
		\end{subfigure}
	\end{minipage}%
	\begin{minipage}{0.5\textwidth}
		\begin{subfigure}{0.9\linewidth}
			\centering
			\caption{\hspace*{11em} \vspace{-0.5em}}
			\hspace{-6cm}
			\includegraphics[width = \textwidth]{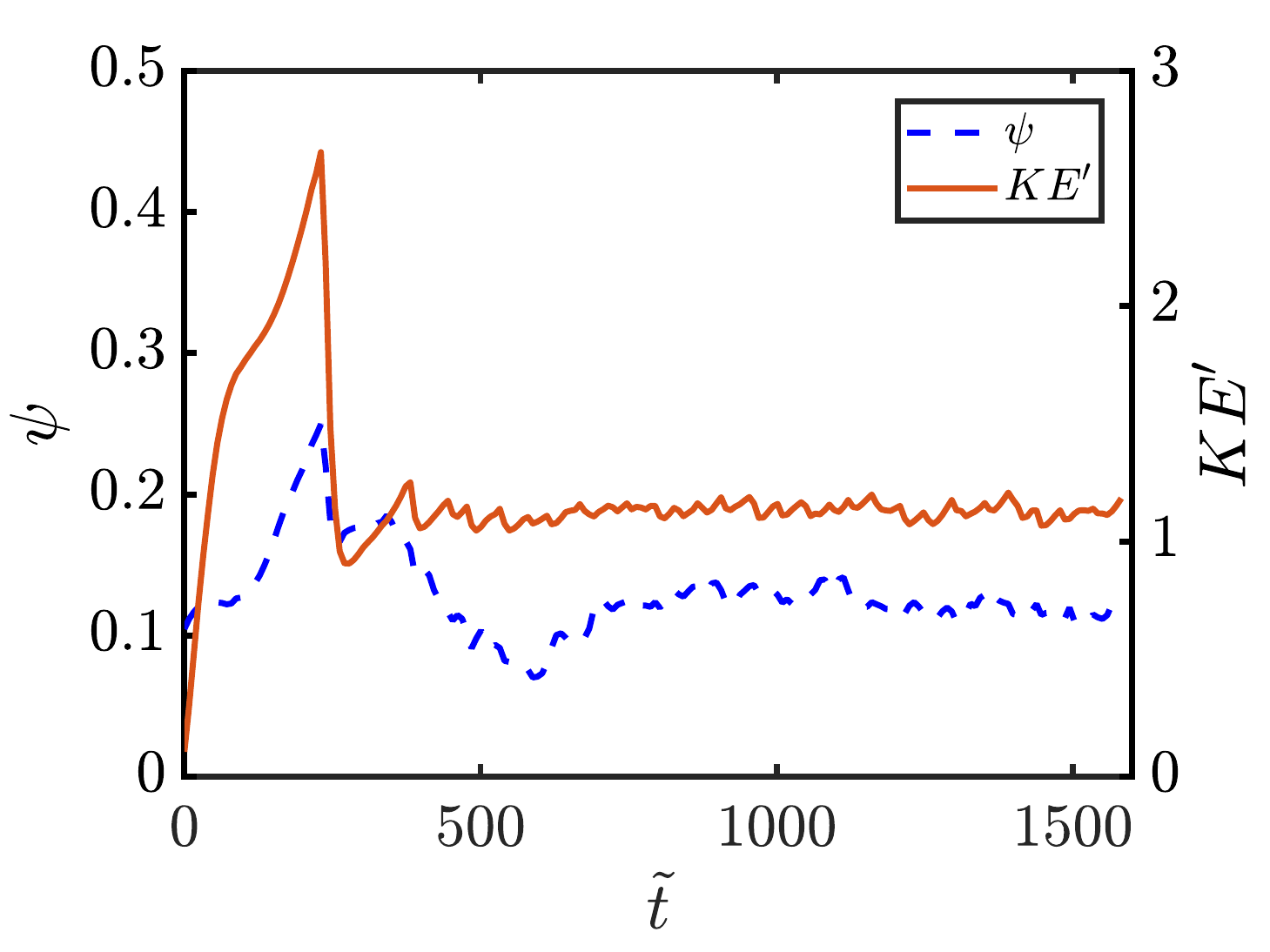}
			\label{fig:fig_2c}
		\end{subfigure}\\[1ex]
		\begin{subfigure}{\linewidth}
			\centering
			\caption{\hspace*{12em} \vspace{-0.5em}}
			\hspace{-4cm}
			\includegraphics[width = 0.8\textwidth]{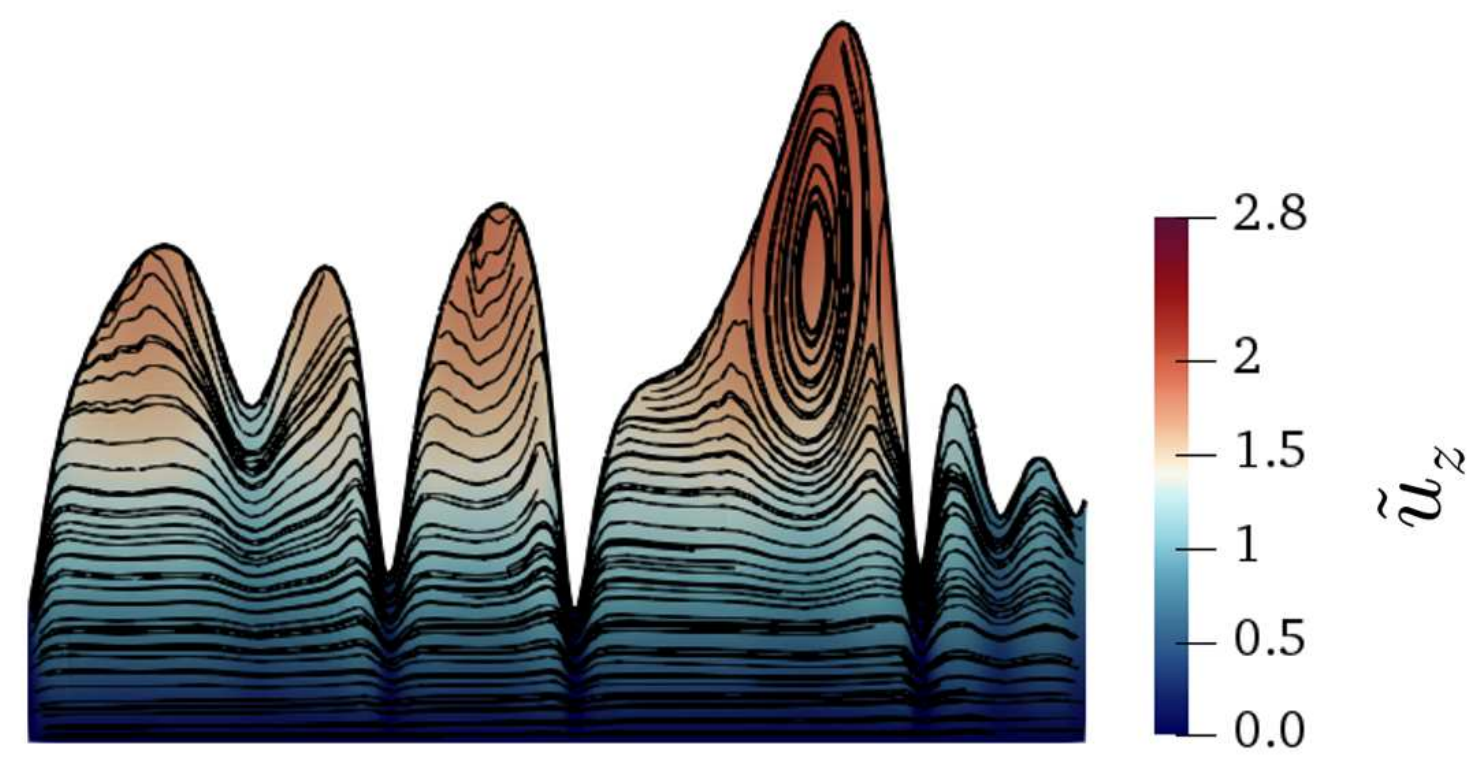}
			\label{fig:fig_2d}
		\end{subfigure}
	\end{minipage}
	\caption{Flow dynamics for the $Ek = 193$ case:  a) time-space plot of interface %
		showing a transition towards a dynamic steady state; b) a snapshot of the film thickness profile at $\tilde{t} =1540$ with the colour bar showing the magnitude of the azimuthal velocity component, $\tilde{u}_\theta$; (c) temporal evolution of the kinetic energy, $KE'$, and film waviness, $\psi$; d) axial velocity component $\tilde{u}_z$ with streamlines in the reference frame of the wave celerity, $c$, for $\tilde{t} = 1540$. The rest of the parameters are $Re = 53$, $Fr = 4.2$, and $We = 0.18$.}
	\label{fig:fig2}
\end{figure}

Fig. \ref{fig:fig2} depicts the flow characteristics of a typical film evolution for $Ek = 193$, which corresponds to an intermediate rotational speed; the rest of the parameter values are fixed at $Re = 53$, $Fr = 4.2$, and $We = 0.18$. Small-amplitude perturbations originating near the domain inlet are amplified downstream under the action of gravity leading to a transition from an essentially waveless to a wavy flow regime. The structure of the emergent waves shown in the time-space plot and the snapshot at $\tilde{t} = 1540$ in Figs. \ref{fig:fig2}a and \ref{fig:fig2}b, respectively, is due to a delicate interplay between gravitational, centrifugal, capillary, inertial, and viscous forces. This structure is characterised by large-amplitude features that interact as they flow in the streamwise direction. It is also clear upon close inspection of Fig. \ref{fig:fig_2b} that the magnitude of the azimuthal velocity component, $\tilde{u}_\theta$, in the film varies over a relatively narrow range, close to the imposed cylinder rotation.  %

\begin{figure}[htbp!]
	\centering
	\begin{subfigure}[b]{0.49\linewidth}
		\centering
		\caption{\hspace*{-2em} \vspace{-0.5em}}
		\includegraphics[width = 0.95\textwidth]{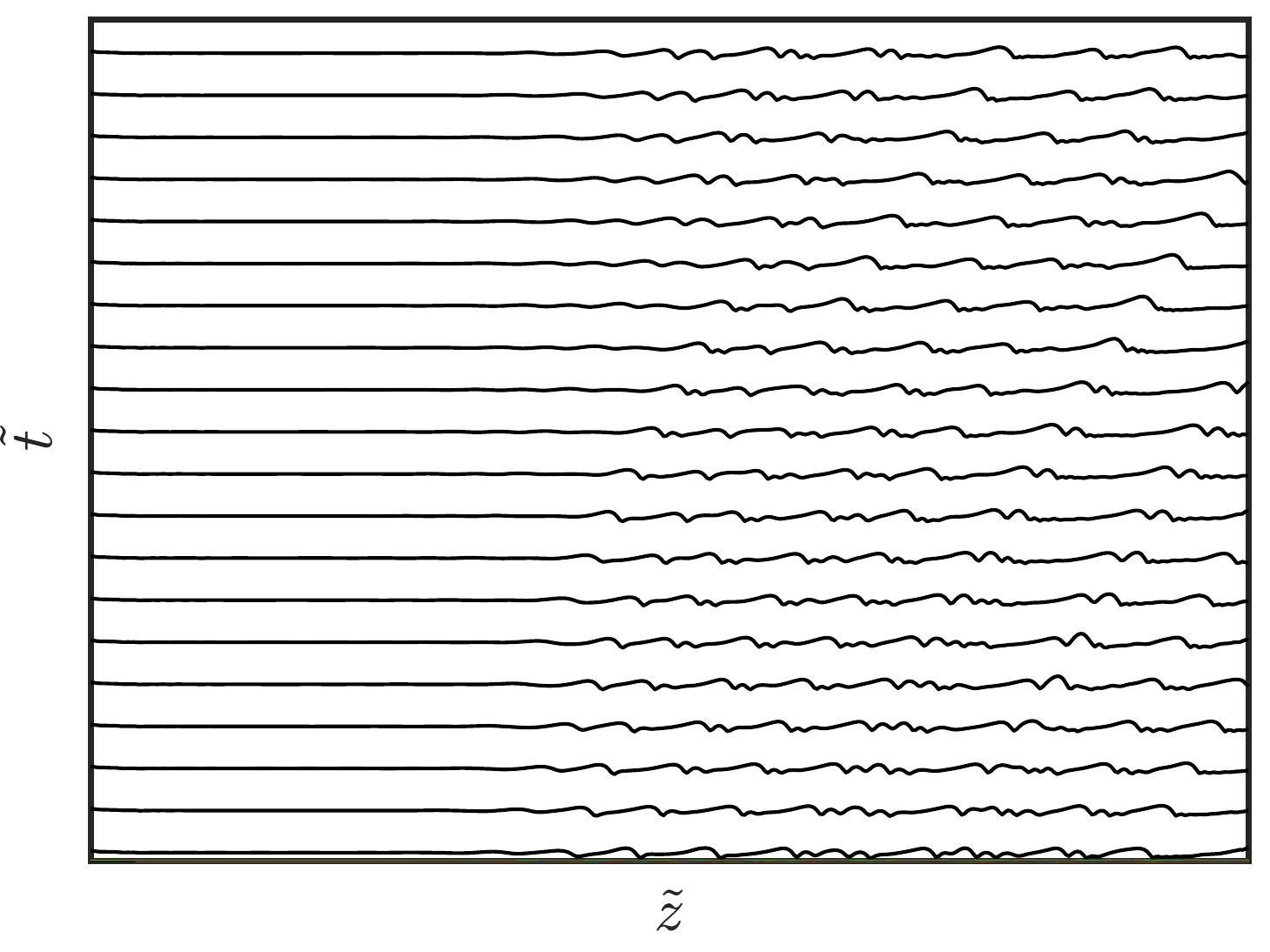}
		\label{fig:fig_3a}\bigskip
	\end{subfigure}
	\begin{subfigure}[b]{0.49\linewidth}
		\centering
		\caption{\hspace*{-2em} \vspace{-0.5em}}
		\includegraphics[width = 0.95\textwidth]{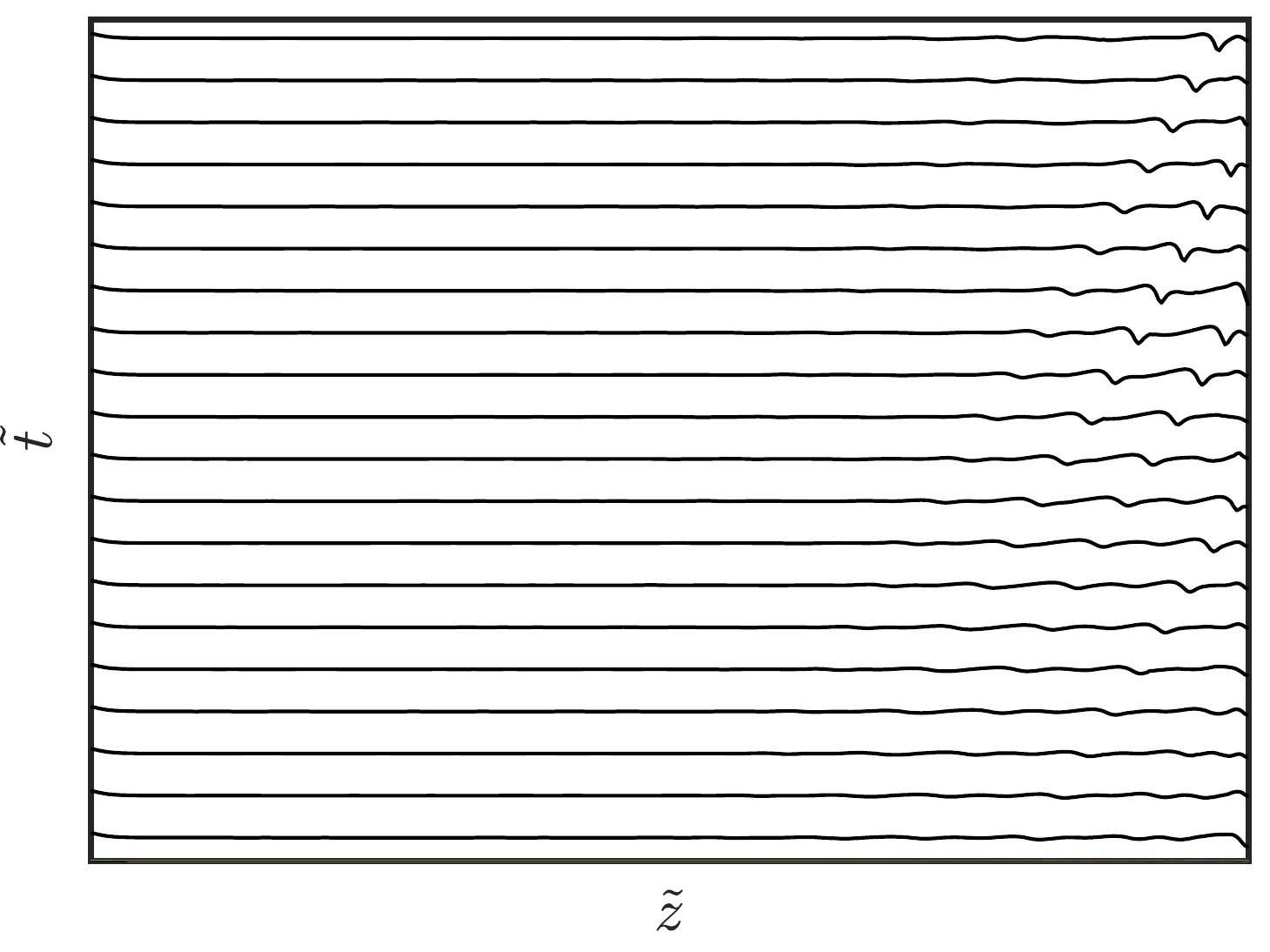}
		\label{fig:fig_3b}\bigskip
	\end{subfigure}
	\begin{subfigure}[b]{0.49\linewidth}
		\centering
		\caption{\hspace*{-3em} \vspace{-0.5em}}
		\includegraphics[width = 0.95\textwidth]{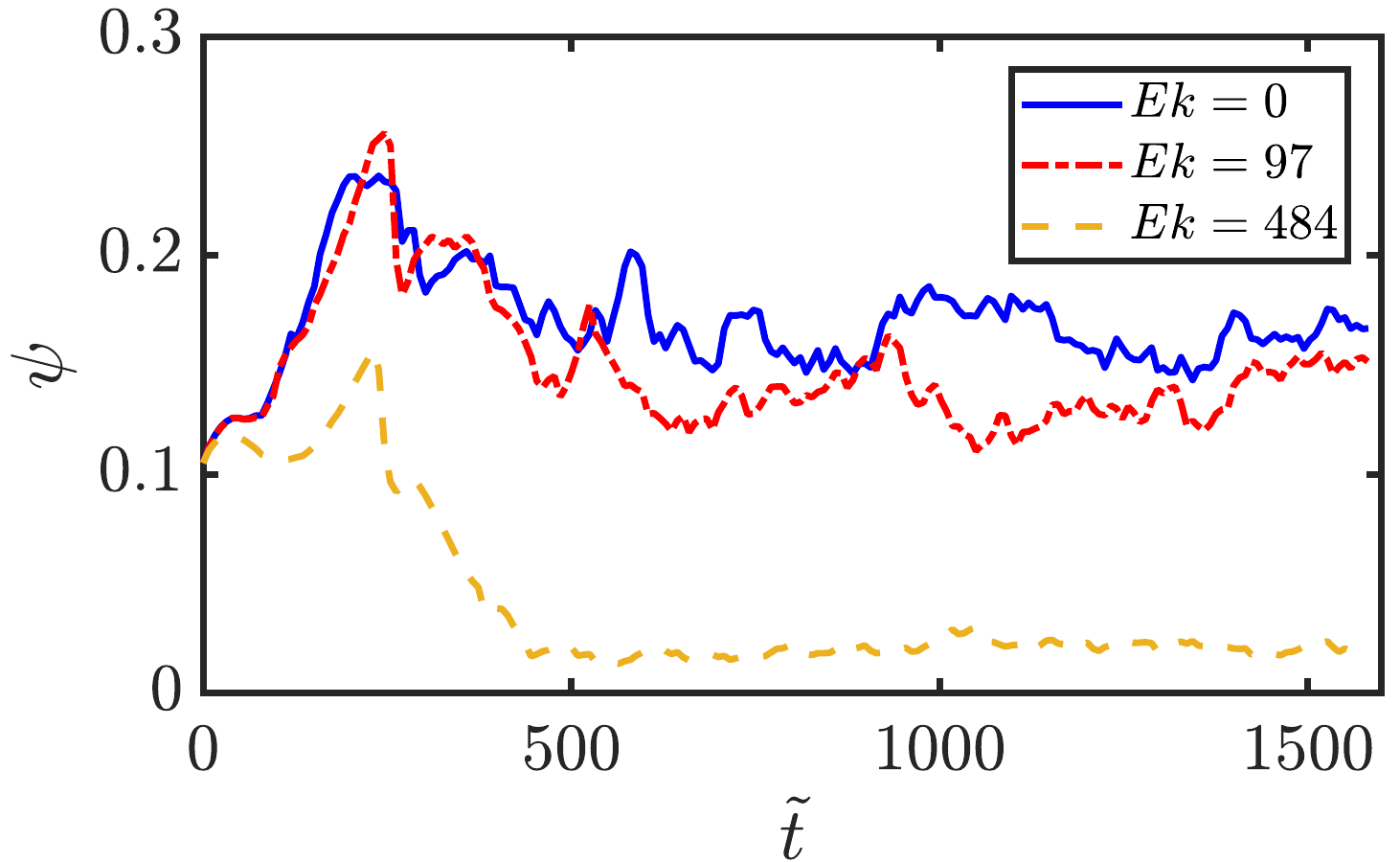}
		\label{fig:fig_3c}
	\end{subfigure}
	\begin{subfigure}[b]{0.49\linewidth}
		\centering
		\caption{\hspace*{-3em} \vspace{-0.5em}}
		\includegraphics[width = 0.95\textwidth]{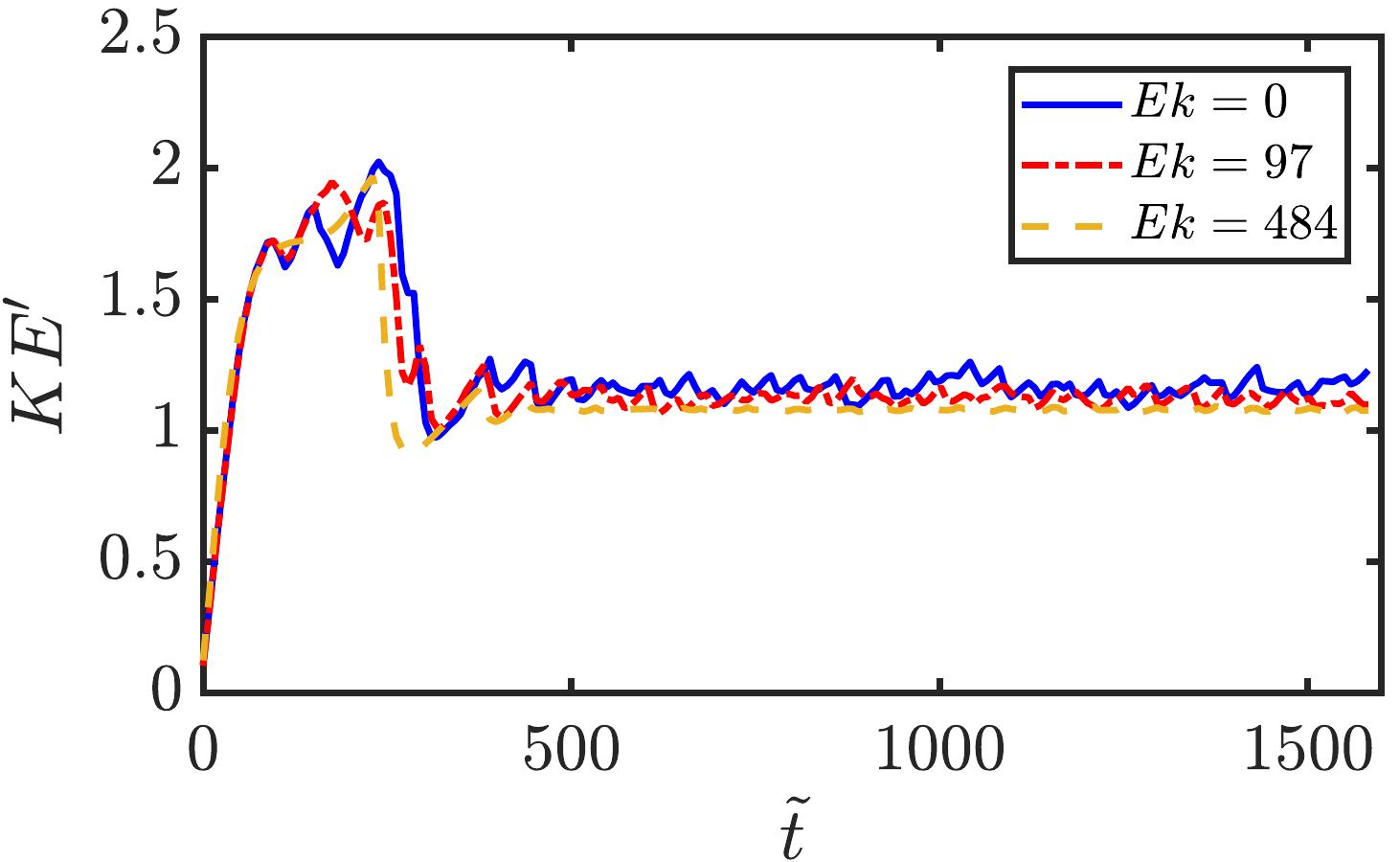}
		\label{fig:fig_3d}
	\end{subfigure}
	\begin{subfigure}{0.55\linewidth}
		\centering
		\caption{\hspace*{-2em} \vspace{-0.5em}}
		\includegraphics[width = \textwidth]{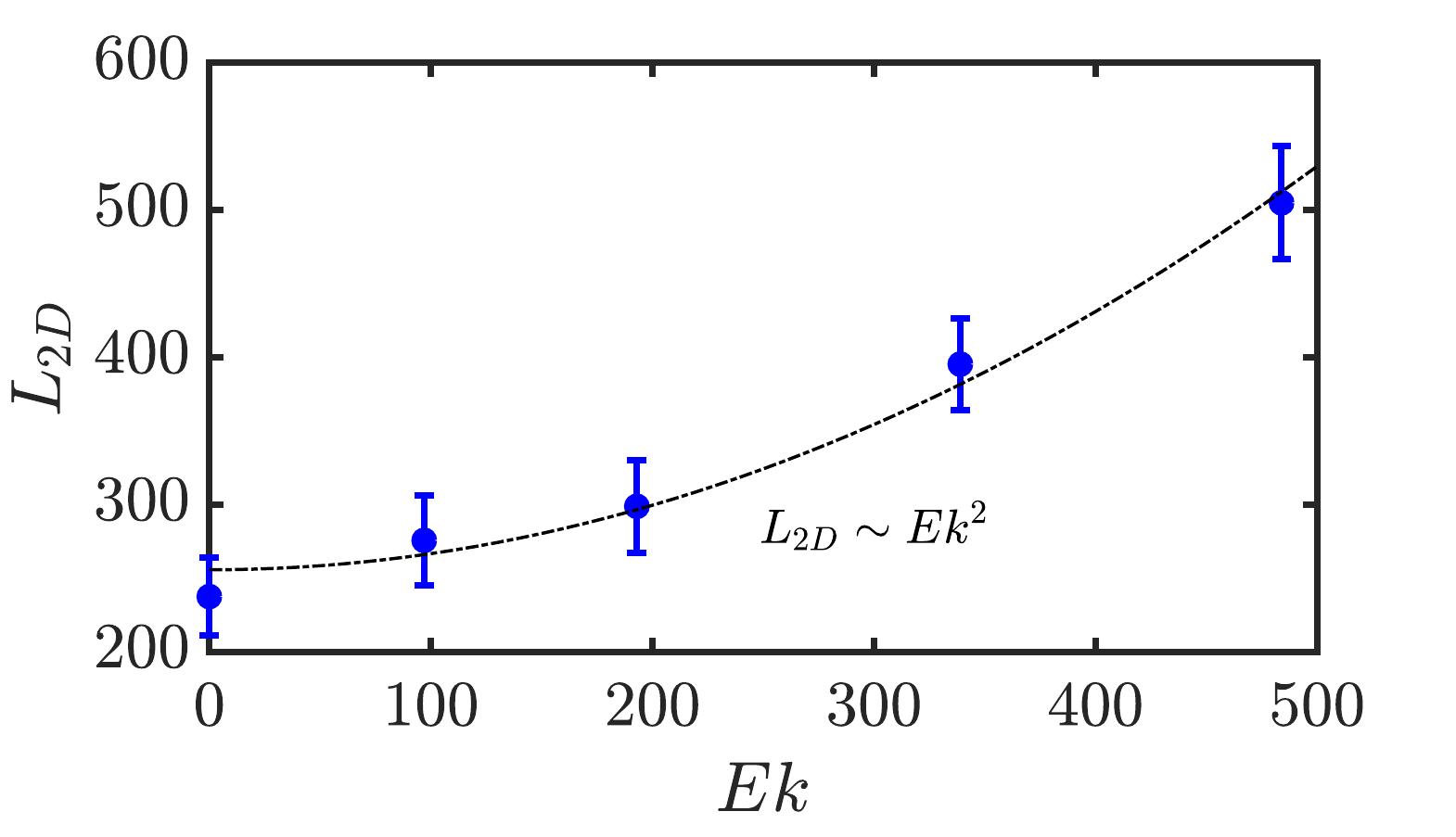}
		\label{fig:fig_3e}
	\end{subfigure}
	\caption{Effect of $Ek$ on the flow: (a) and (b) time-space plots of the interface for  $Ek = 0$ and 484, respectively; (c) and (d) temporal evolution of the film waviness, $\psi$, and kinetic energy, $KE'$, respectively, for $Ek=0$, 97, and 484; (e) variation of the domain length beyond which a transition to 3D structures is observed, $L_{2D}$, with $Ek$. The rest of the parameter values remain unaltered from Fig. \ref{fig:fig2}.}
	\label{fig:fig3}
\end{figure}

In Fig. \ref{fig:fig_2c} we track the kinetic energy, $KE'$, and the film waviness, $\psi$, respectively given by 
\begin{equation}\label{KEt}
KE' = KE - KE_\theta,
\end{equation}
\begin{equation}\label{psi1}
\psi=\int^{\tilde{L}_z}_0|\tilde{h}-1| d \tilde{z},
\end{equation}
where $KE'=\int^{\tilde{L}_z}_0\int^{\tilde{L}_h}_0 \tilde{r} |\tilde{\mathbf{{u^{\prime}}}}|^2d\tilde{r}d\tilde{z}$ is the kinetic energy %
in a rotating frame of reference.  %
The film waviness $\psi$ is defined as a measure of the fluctuations of the film from the Nusselt film height. Fig. \ref{fig:fig_2c}  shows clearly the point at which a dynamic steady-state is reached %
after approximately $\tilde{t} = 600$ (though there remain low-amplitude fluctuations in $\psi$ past this point in time). %

\begin{figure} 
	\centering
	\begin{subfigure}[t]{0.49\linewidth}
		\centering
		\caption{\hspace*{-2em}}
		\includegraphics[width = 0.9\textwidth]{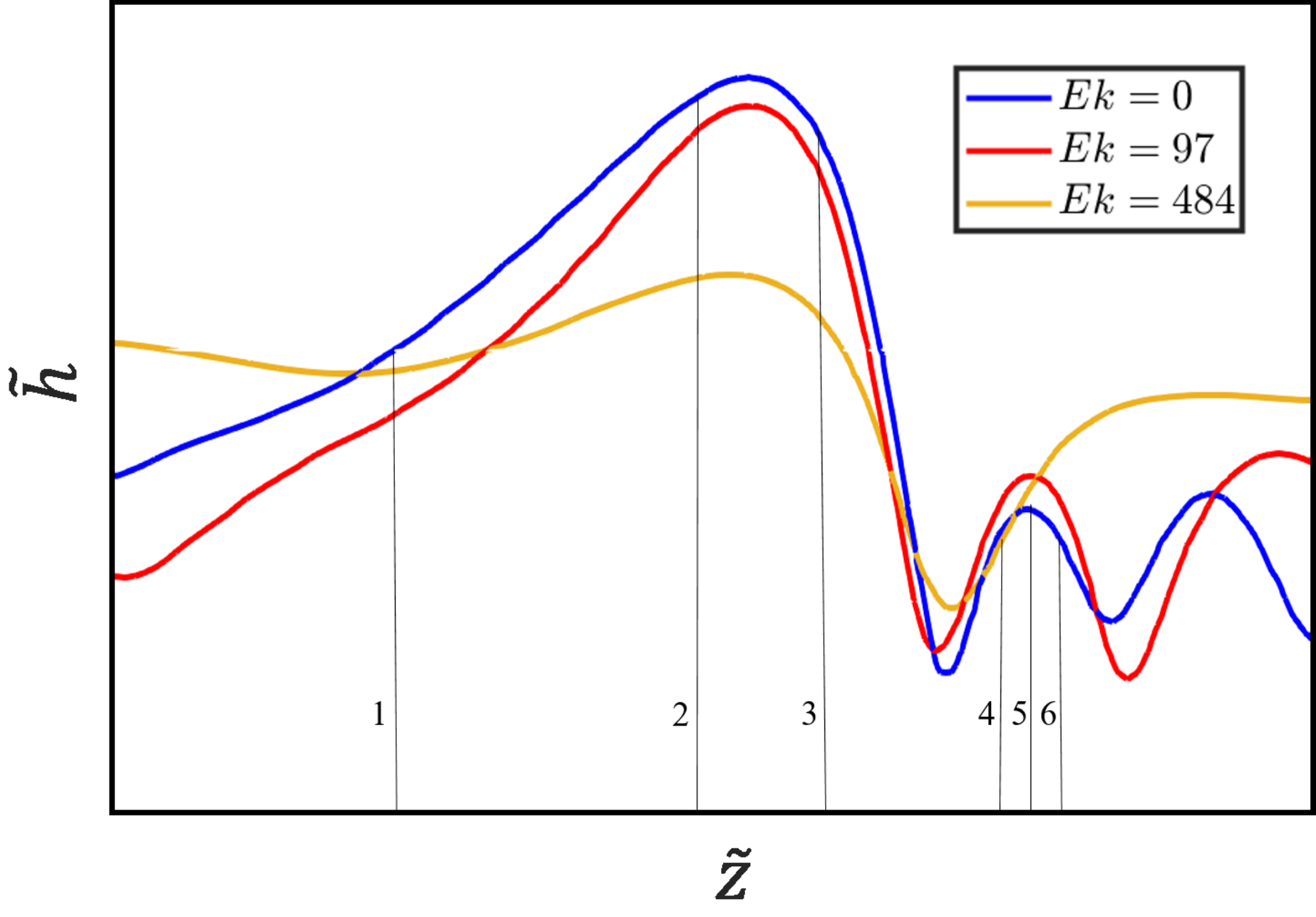}
		\hspace{-1.5em}
		\label{fig:fig_3a1}
	\end{subfigure}
	\begin{subfigure}[t]{0.49\linewidth}
		\centering
		\caption{\hspace*{-2em} \vspace{-0.5em}}
		\includegraphics[width = 0.95\textwidth]{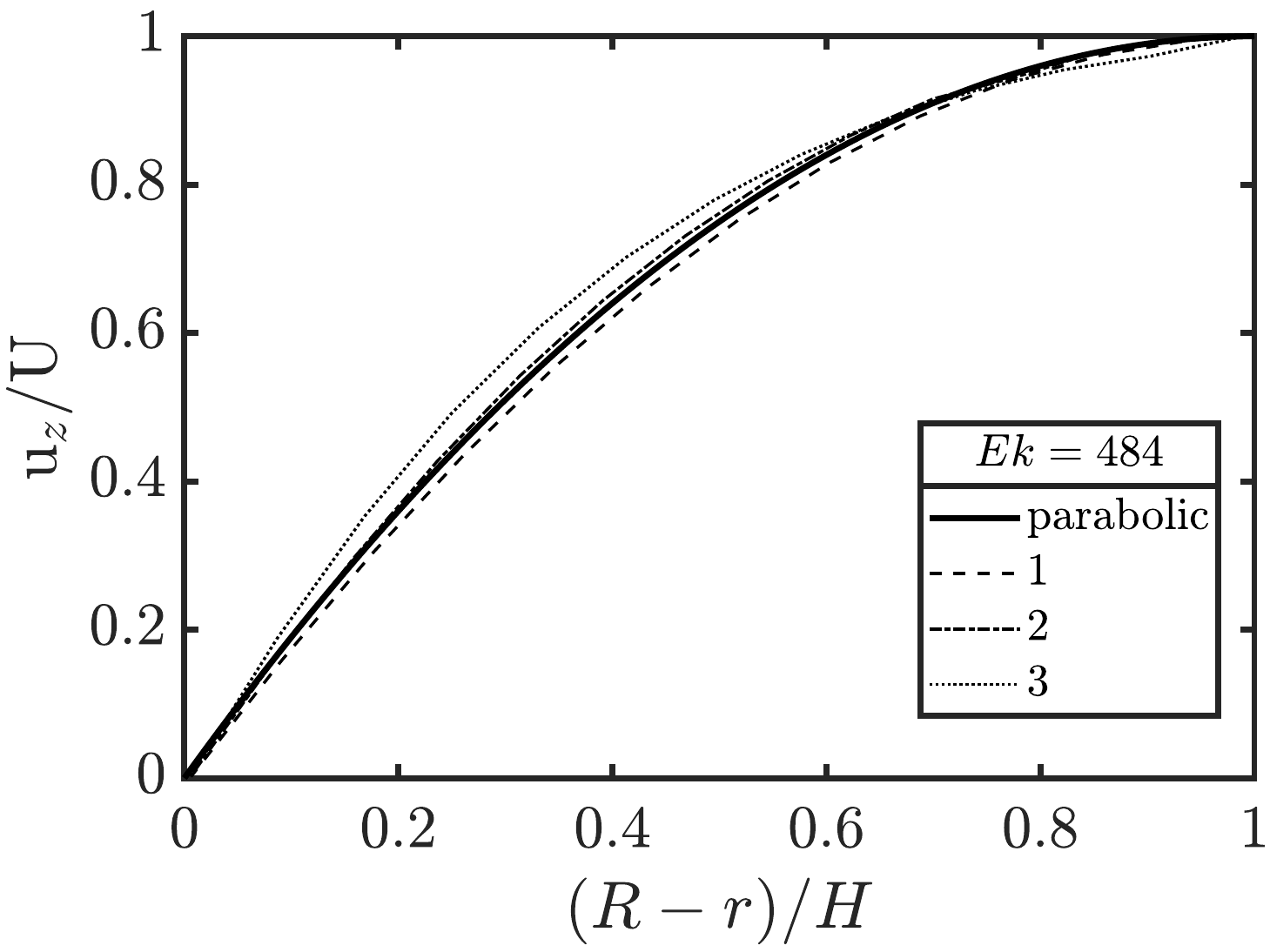}
		\label{fig:fig_3b1}
	\end{subfigure}
	\begin{subfigure}[b]{0.49\linewidth}
		\centering
		\caption{\hspace*{-3em} \vspace{-0.5em}}
		\includegraphics[width = 0.95\textwidth]{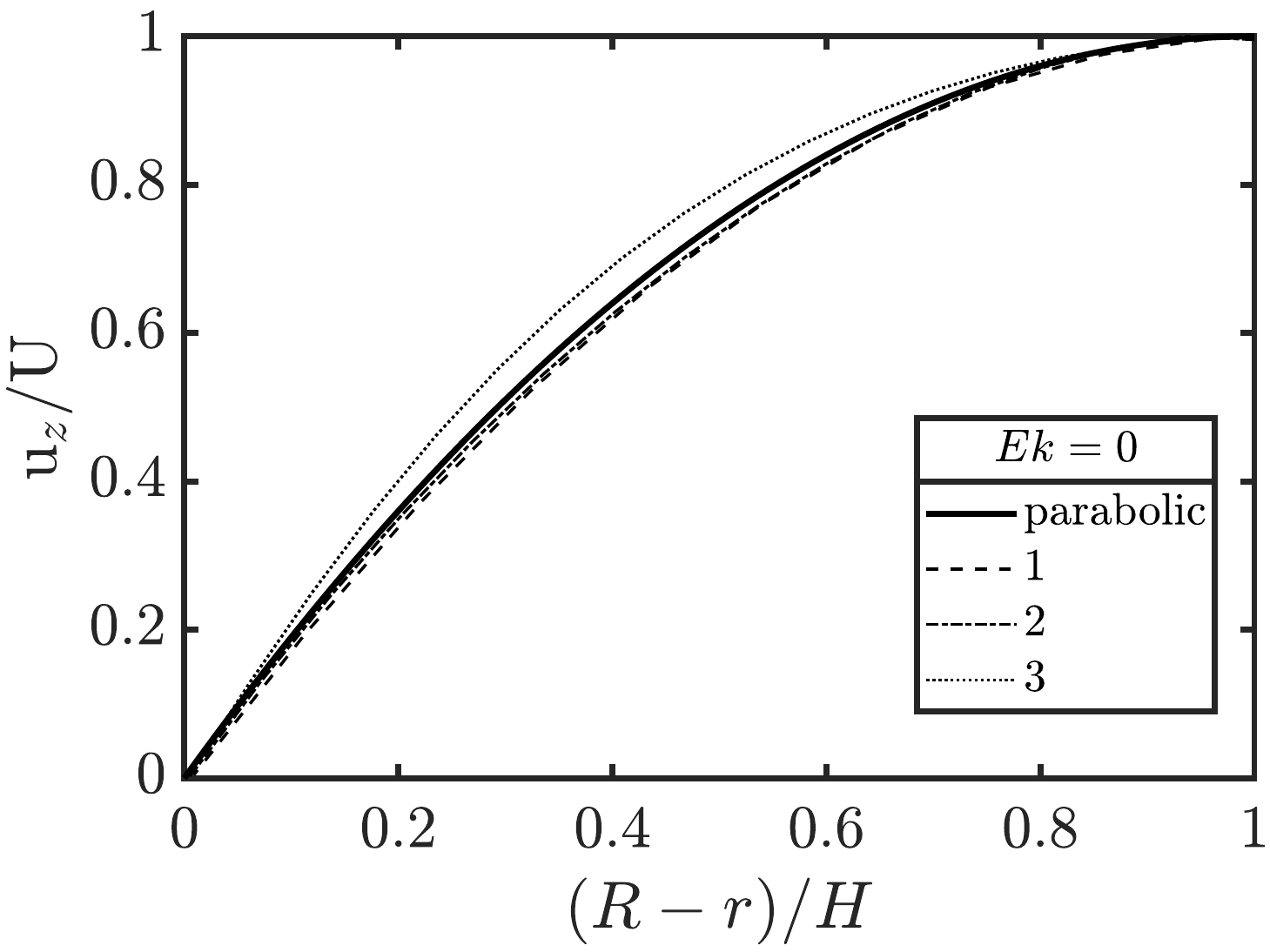}
		\label{fig:fig_3c1}
	\end{subfigure}
	\begin{subfigure}[b]{0.49\linewidth}
		\centering
		\caption{\hspace*{-3em} \vspace{-0.5em}}
		\includegraphics[width = 0.95\textwidth]{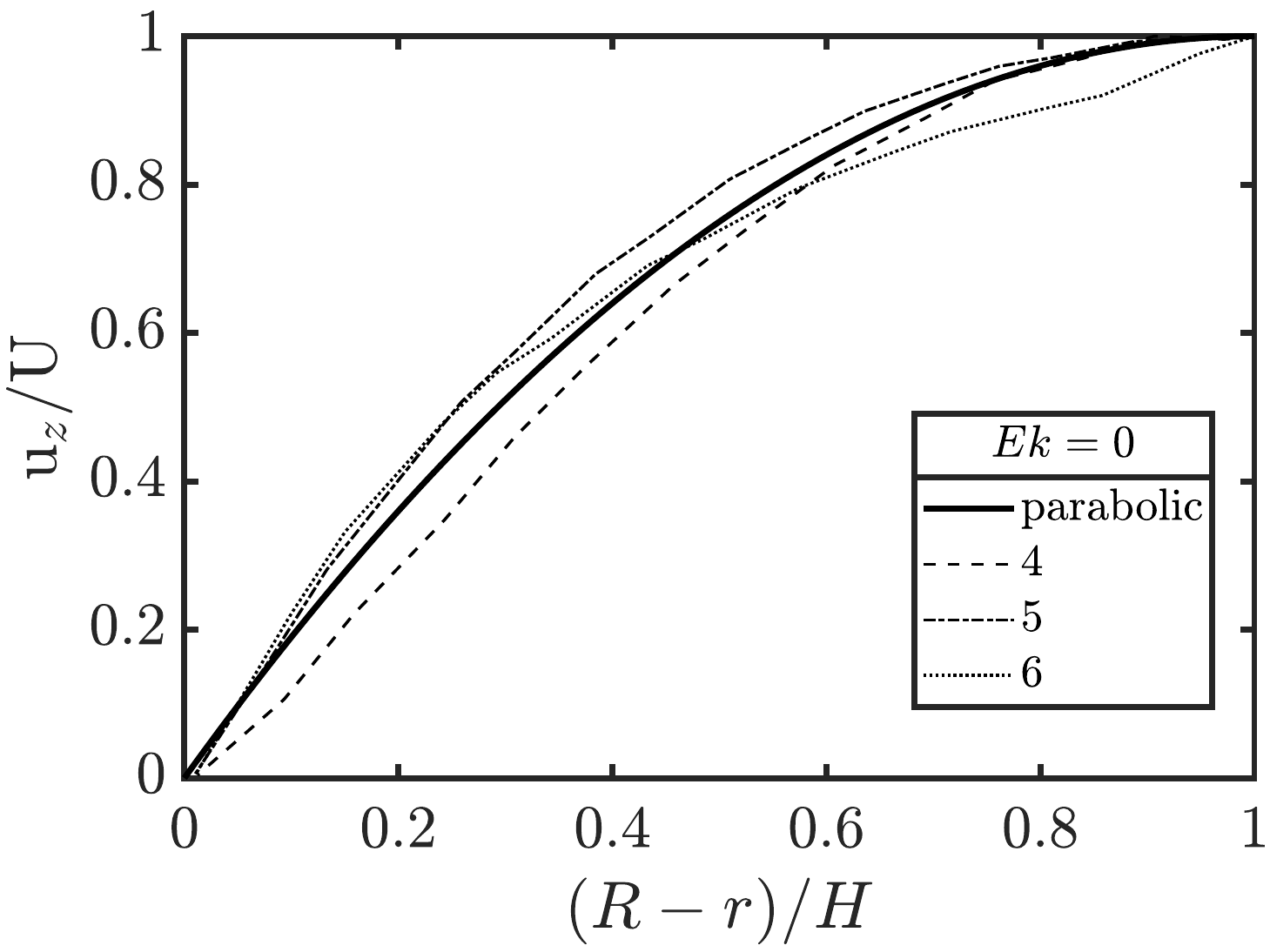}
		\label{fig:fig_3d1}
	\end{subfigure}
	\begin{subfigure}{0.49\linewidth}
		\centering
		\caption{\hspace*{-2em} \vspace{-0.5em}}
		\includegraphics[width = 0.95\textwidth]{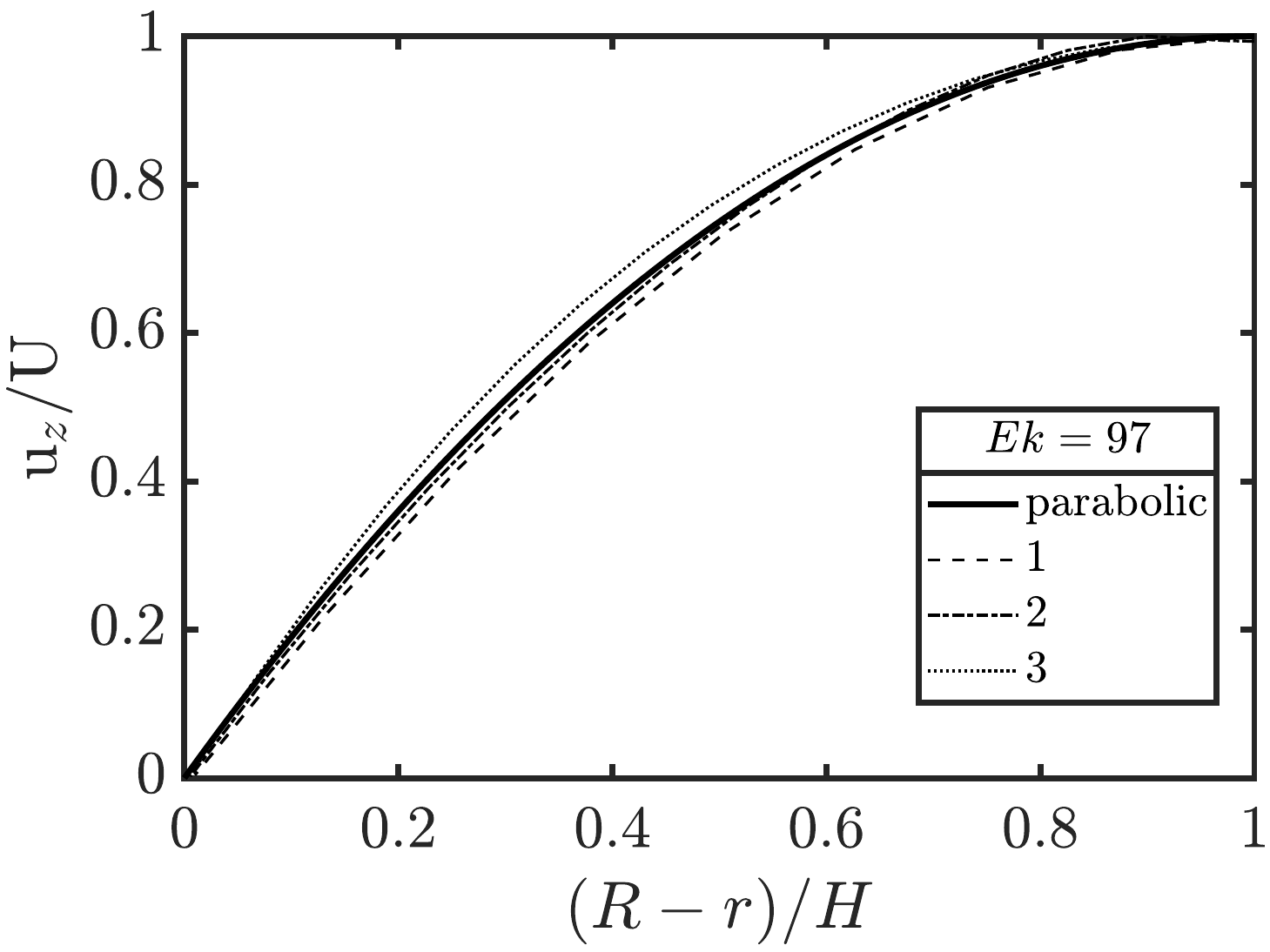}
		\label{fig:fig_3e1}
	\end{subfigure}
	\begin{subfigure}{0.49\linewidth}
		\centering
		\caption{\hspace*{-2em} \vspace{-0.5em}}
		\includegraphics[width = 0.95\textwidth]{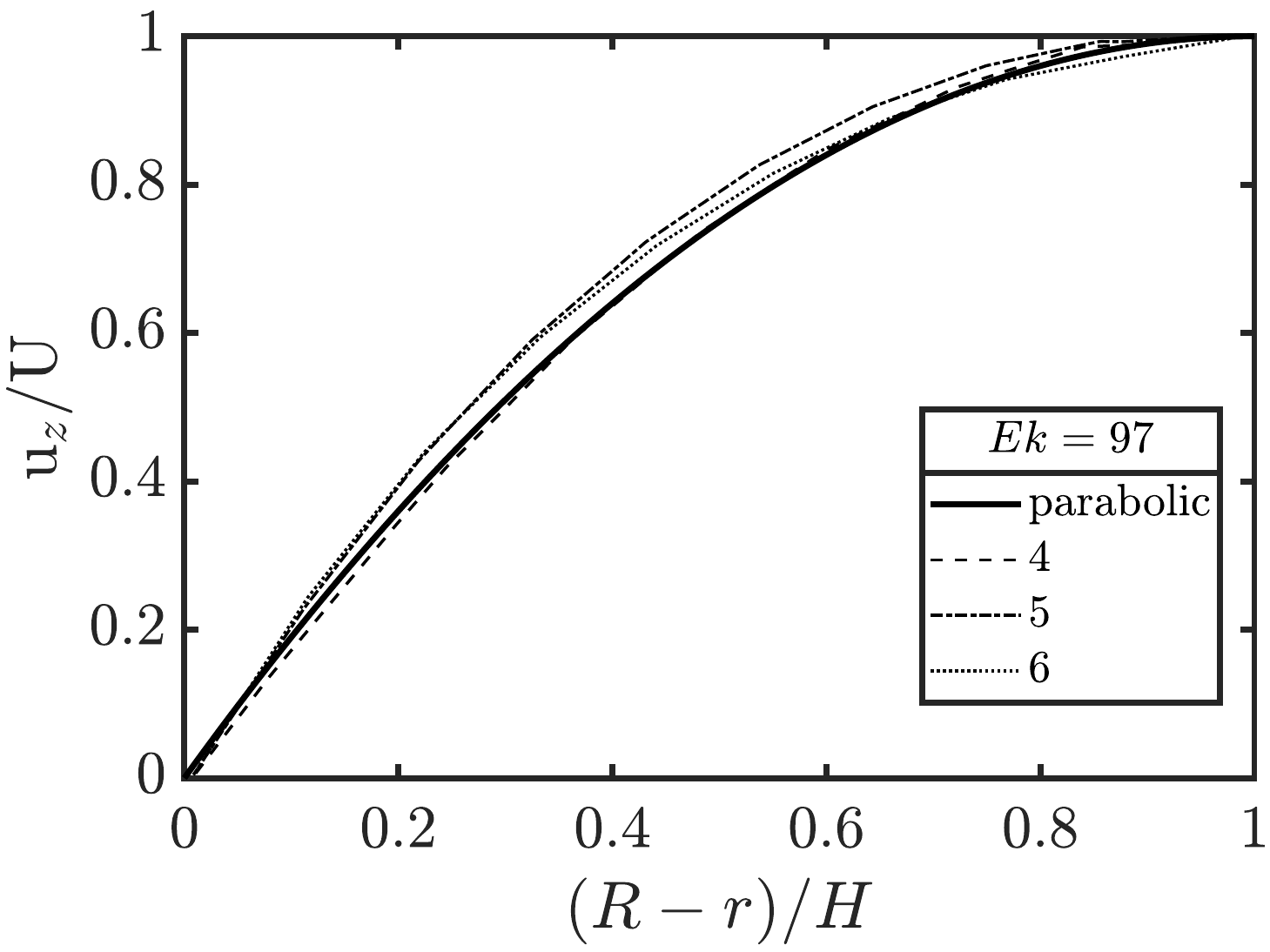}
		\label{fig:fig_3f1}
	\end{subfigure}
	\caption{Velocity profiles in the film underneath the the solitary waves for various $Ek$ with the rest of the parameters remaining unaltered from Fig. \ref{fig:fig2}: (a) three interfacial profiles for $Ek=0$, 97, and 484, with the points `1'-`6' marking the locations at which the radial variation of the axial velocity component normalised by its maximal value, $u_z/U$, is plotted in (b), (c) and (d), and (e) and (f) for $Ek=484$, 0, and 97, respectively; also shown in (b)-(f) is a parabolic profile for reference. The radial coordinate is normalised by the local maximum film height $H$}
	\label{fig:fig4}
\end{figure}

The time-space plot depicted in Fig. \ref{fig:fig_2a} can be used to estimate the wave speed or celerity, $c$, by tracking a single wave in the domain and noting the distance moved $\Delta z$ in time $\Delta t$. %
This was found to be approximately 0.4 m.s$^{-1}$ essentially independent of $Ek$ and, in turn, %
of the rotational speed of the cylinder, which acts in the azimuthal direction; it is primarily influenced by the constant gravitational force %
in the axial direction. %
In some cases, the local wave speed can exceed $c$ leading to the formation of %
a recirculation zone %
in the reference frame of the wave celerity \cite{denner_jfm_2018,rohlfs_jfm_2015}, which gives rise to enhanced mixing. %
These flow characteristics are less prevalent at high $Ek$ as the impact of rotation mitigates the formation of large, fast waves due to the increased stabilisation. %
This is industrially relevant, as operating at higher rotational speeds may be preferential for certain applications for transport control, for instance, but will demote mixing within the waves. %
Streamlines in the reference frame of the wave celerity are plotted in Fig. \ref{fig:fig_2d} where the recirculation zone is clear in the larger, faster-moving wave but absent in the smaller one.

Having analysed the flow for a single intermediate $Ek$ value, Fig. \ref{fig:fig3} shows the effect on the dynamics across a range of $Ek$ with all other parameters kept constant at $Re=53$, $Fr=4.2$, $We=0.18$, and $\beta=6.9\times 10^{-3}$. Time-space plots of the dynamic steady-state are shown for $Ek = 0$ in Fig \ref{fig:fig_3a} and $Ek = 339$ in Fig \ref{fig:fig_3b}, and whilst the natural evolution from a waveless to a wavy regime is notable in both cases, there are distinct differences. Qualitatively the reduction in the degree of waviness is apparent %
due to the stabilising effect. %
There is also a clear extension in the waveless regime with an increase in $Ek$, which can be quantified by estimating the transition from the waveless to a wavy regime, beyond a length $L_{2D}$ computed  %
according to $\tilde{z}=L_{2 D}, \text { if } \tilde{h}<0.98 \text { or } \tilde{h}>1.02$. %
As shown qualitatively in Figs. \ref{fig:fig_3a} and \ref{fig:fig_3b}, this entry length, establishing the onset of 2D waves, increases monotonically with %
$Ek$. One expects $L_{2D}$ to have the following dependence on the operating parameters:
\begin{equation}
\frac{L_{2D}}{h_N} \sim \frac{\Omega^2 R}{g}.
\end{equation}
From the definition of $Ek$, $Re$, $h_N$, and $\beta$, this relation can be re-expressed as follows
\begin{equation}
\frac{L_{2D}}{h_N} \sim \beta \frac{Ek^2}{Re},    
\end{equation}
thus $L_{2D} \sim Ek^2$ for fixed $\beta$ and $Re$.
The dynamic evolution of $KE'$ and $\psi$ for a range of $Ek$ is also shown in Figs. \ref{fig:fig_3c} and \ref{fig:fig_3d}, respectively. It is seen clearly that variation of $Ek$ has a profound effect on the film waviness: we observe that there is a decrease in the film waviness with an increase in $Ek$. At the largest $Ek$ examined, the film waviness is suppressed significantly, with $\psi$ approaching a steady value close to zero, compared with lower $Ek$. This is as expected as an increase in $Ek$ is tantamount to an increase in the rotational speed, which  increases the centrifugal force, stabilising the flow. In contrast, the temporal variation of $KE'$ is weakly-dependent on $Ek$ though it provides an indication of when a dynamic steady-state is reached (beyond $\tilde{t} \approx 500$). 
In Fig. \ref{fig:fig_3a1}, we show a comparison of waves formed at different $Ek$ with the rest of the parameters kept unaltered from Fig. \ref{fig:fig3}; each one of these waves corresponds to a travelling-wave extracted from the domain at distances beyond $L_{2D}$. The first observation that can be made is the decrease in the amplitude of the wave peaks with an increase in $Ek$, due to the increasing stabilising force. It is also seen that at high $Ek$, the wave structure is altered significantly with a severe depression in the peak, and a much less pronounced distinction between the primary wave and the capillary waves downstream, which is characteristic of solitary waves in falling film flows. It should be noted that the wave depicted for $Ek=484$ in Fig. \ref{fig:fig_3a1}, however, does not belong to the nearly-sinusoidal wave family but a solitary one. 
Indeed, we have found that the centrifugal forces arising from the imposed rotation simply delay the emergence of the solitary waves, which is in accordance with the observation made in Fig. \ref{fig:fig_3e} that illustrates the monotonic growth of $L_{2D}$ with $Ek$. Further investigation into the nature of fully-developed waves is conducted below. %

Due to the complex nature of the waves, one would expect the %
velocity profiles within the thin films to be %
non-parabolic \cite{malamataris_pf_2002}. %
We show in Figs. \ref{fig:fig_3b1}-\ref{fig:fig_3f1} the shape of these profiles %
at the designated points in Fig. \ref{fig:fig_3a1}. %
Inspection of %
Fig. \ref{fig:fig_3b1} reveals that the velocity profiles are %
close to parabolic upstream of the wave peaks, in the region in the which a flat film is approached. %
This is in contrast to Fig. \ref{fig:fig_3c1}, where profile `3' specifically bulges over the parabolic curve. This feature still exists for $Ek = 97$ in Fig. \ref{fig:fig_3e1}, though to a lesser extent, reflecting the stabilising effect of cylinder rotation that promotes %
more parabolic-type profiles. 
This trend is also apparent upon inspection of the profiles within the capillary waves, shown in Figs. \ref{fig:fig_3d1} and \ref{fig:fig_3f1}, %
with the profiles tending towards the parabolic reference curve with increasing $Ek$. %
The profiles can also be used to determine the extent of recirculation within the wave. In Figs. \ref{fig:fig_3c1} and \ref{fig:fig_3e1}, profiles `2' and `3' bulge below and above the parabolic profile, respectively, and cross one another, indicative of a recirculation zone within the wave, a common feature of solitary wave profiles \cite{gao_jcp_2003}. %

\begin{figure}
	\centering
	\begin{subfigure}[b]{0.49\linewidth}
		\centering
		\caption{\hspace*{-1.5em} \vspace{-0.5em}}
		\includegraphics[width = \textwidth]{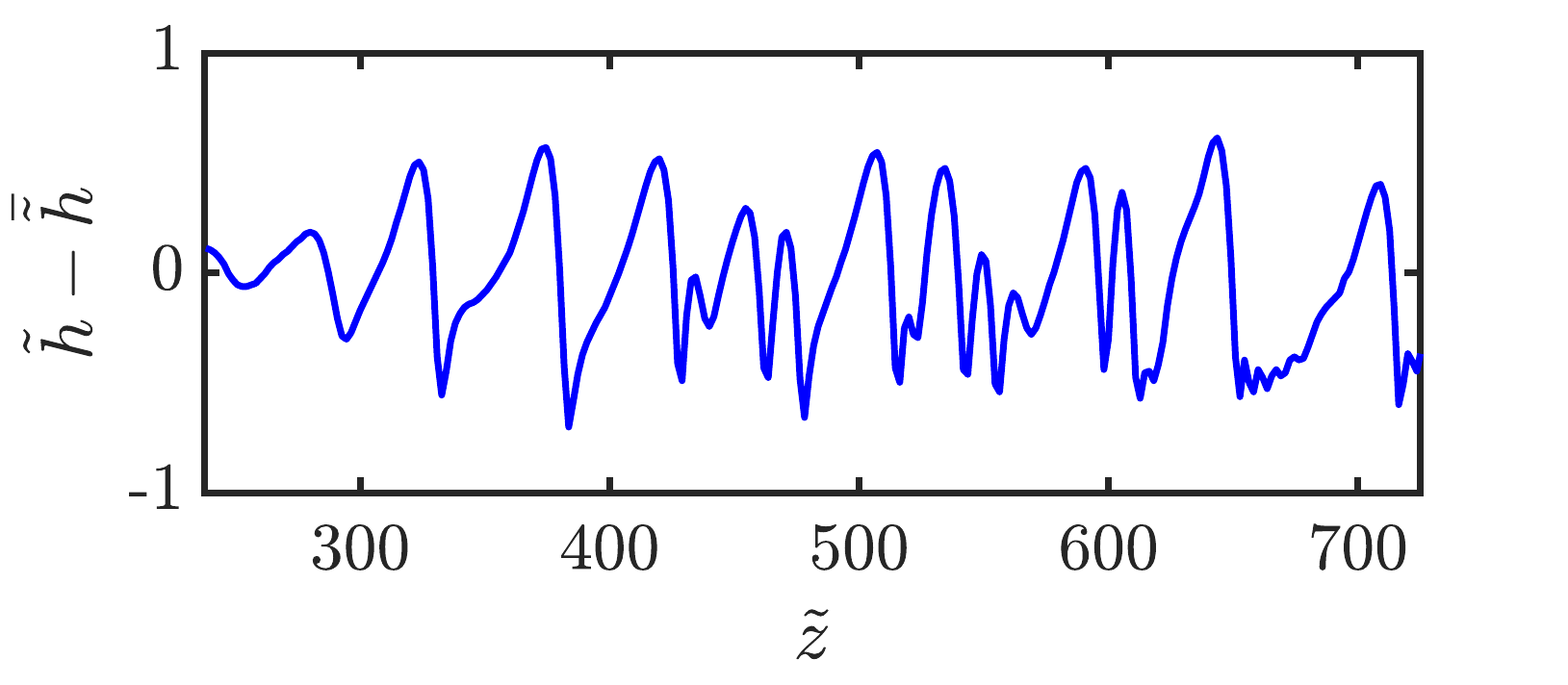}
		\label{fig:fig_5a}
	\end{subfigure}
	\begin{subfigure}[b]{0.49\linewidth}
		\centering
		\caption{\hspace*{-1.5em} \vspace{-0.5em}}
		\includegraphics[width = \textwidth]{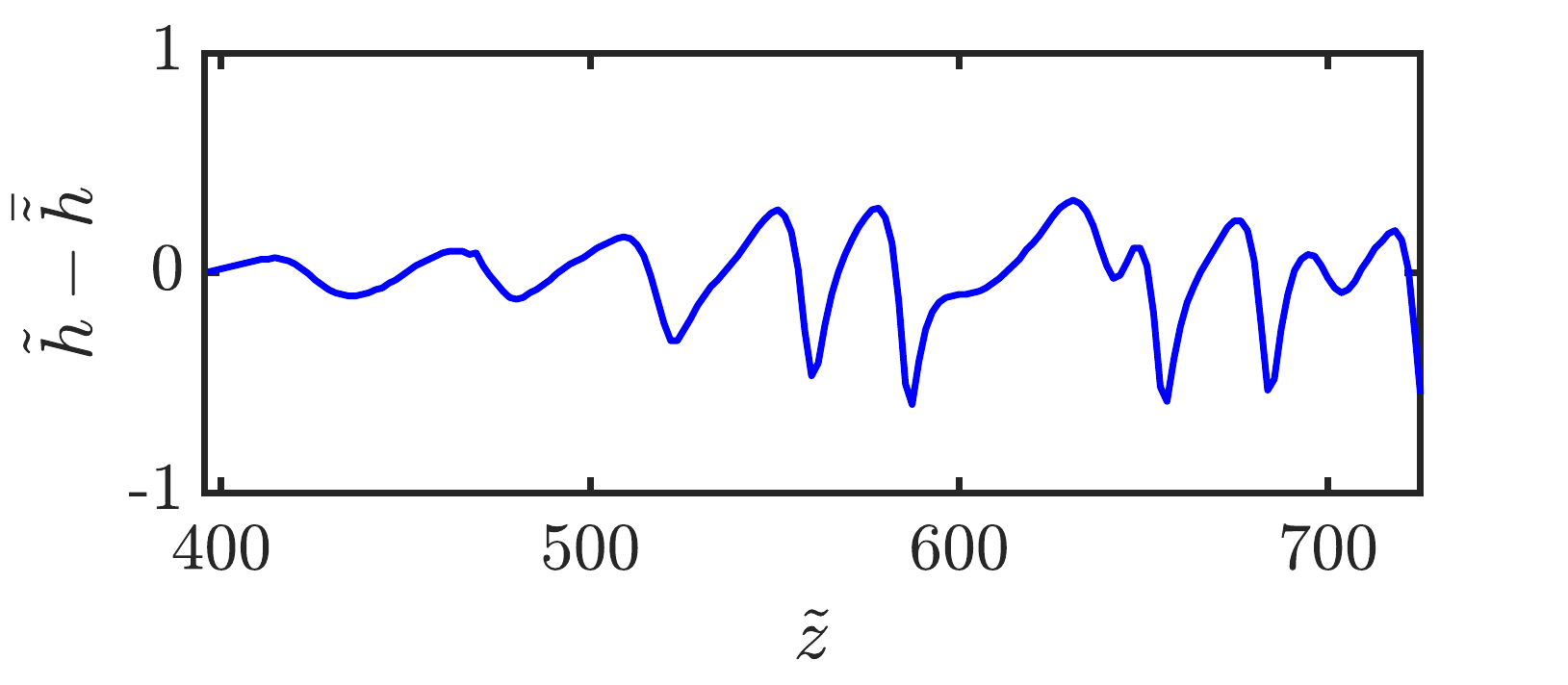}
		\label{fig:fig_5b}
	\end{subfigure}
	\begin{subfigure}{0.49\linewidth}
		\centering
		\caption{\hspace*{-1.5em} \vspace{-0.5em}}
		\includegraphics[width = \textwidth]{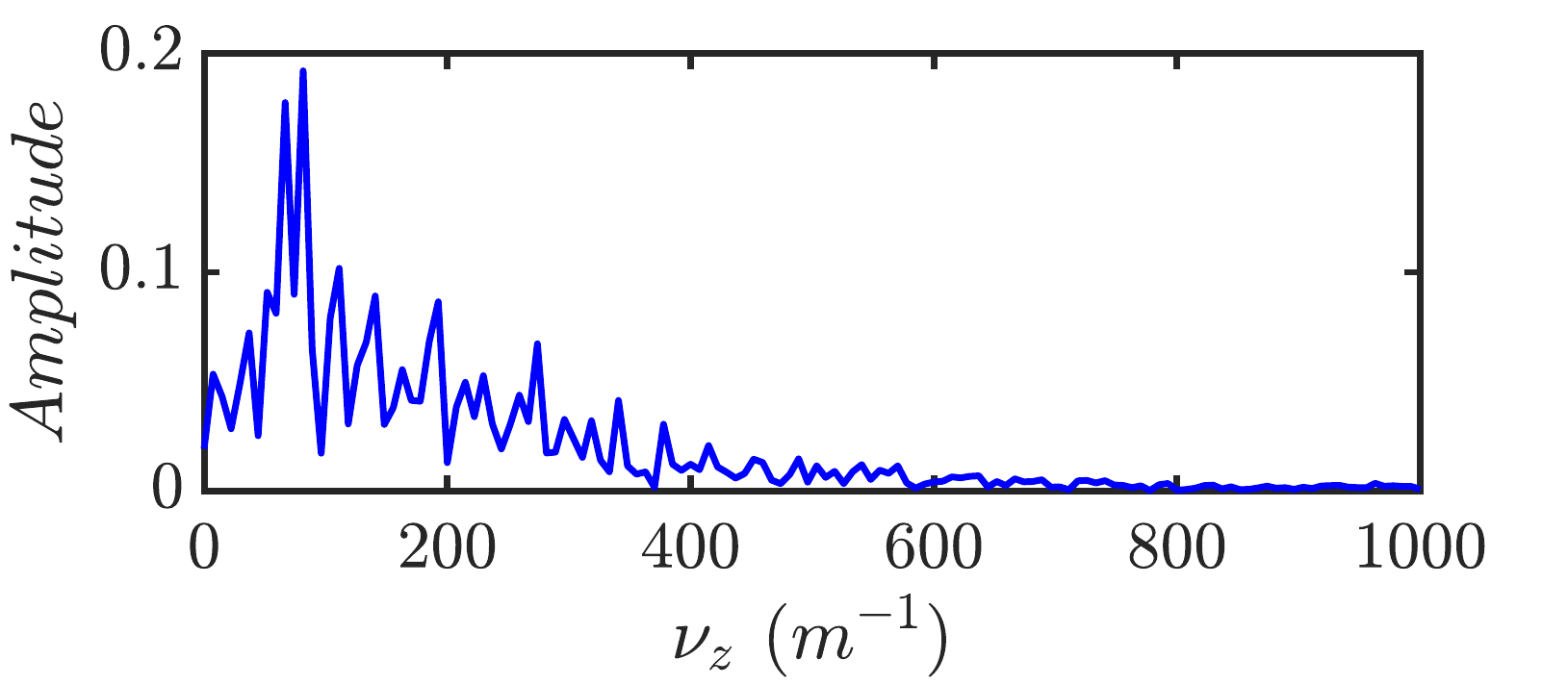}
		\label{fig:fig_5c}
	\end{subfigure}
    \begin{subfigure}{0.49\linewidth}
    	\centering
    	\caption{\hspace*{-1.5em} \vspace{-0.5em}}
    	\includegraphics[width = \textwidth]{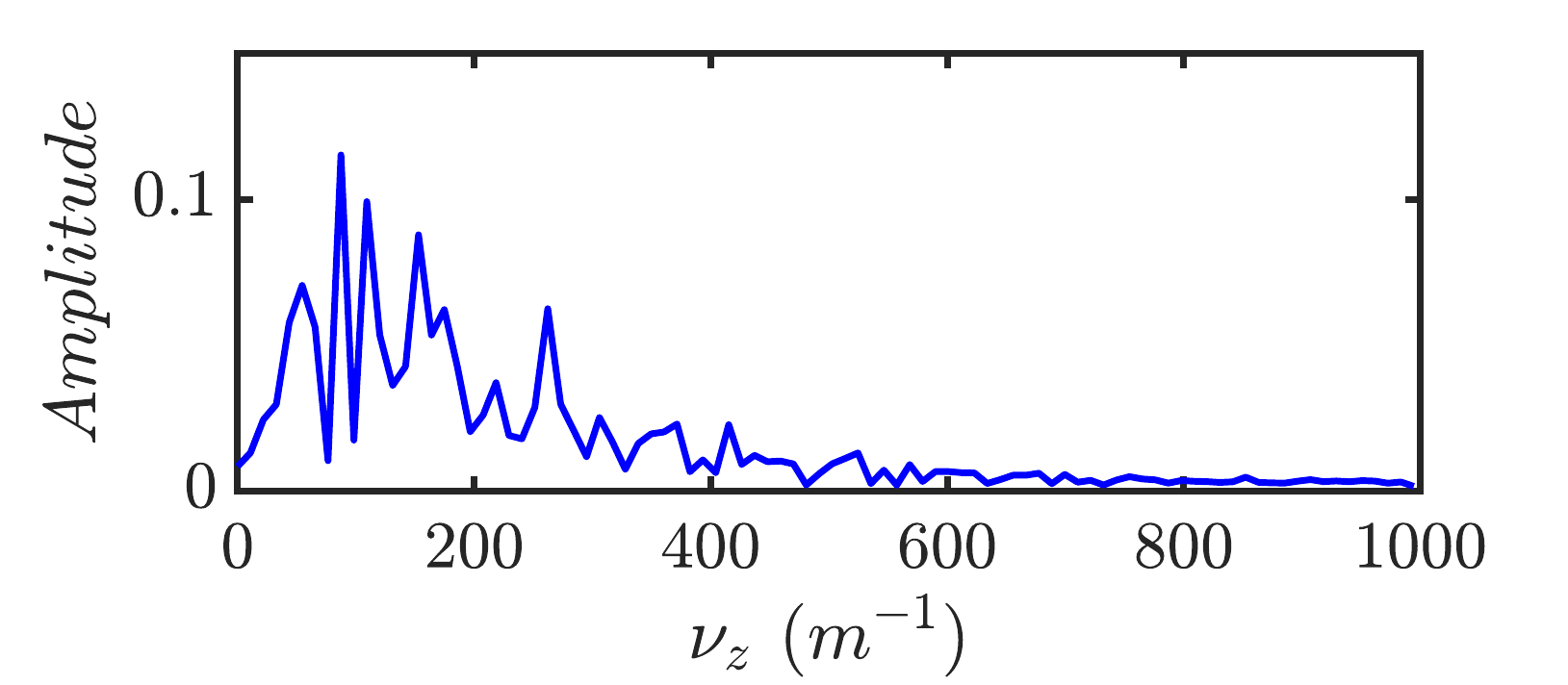}
    	\label{fig:fig_5d}
    \end{subfigure}
	\begin{subfigure}{0.49\linewidth}
	\centering
	\caption{\hspace*{-1.5em} \vspace{-0.5em}}
	\includegraphics[width = \textwidth]{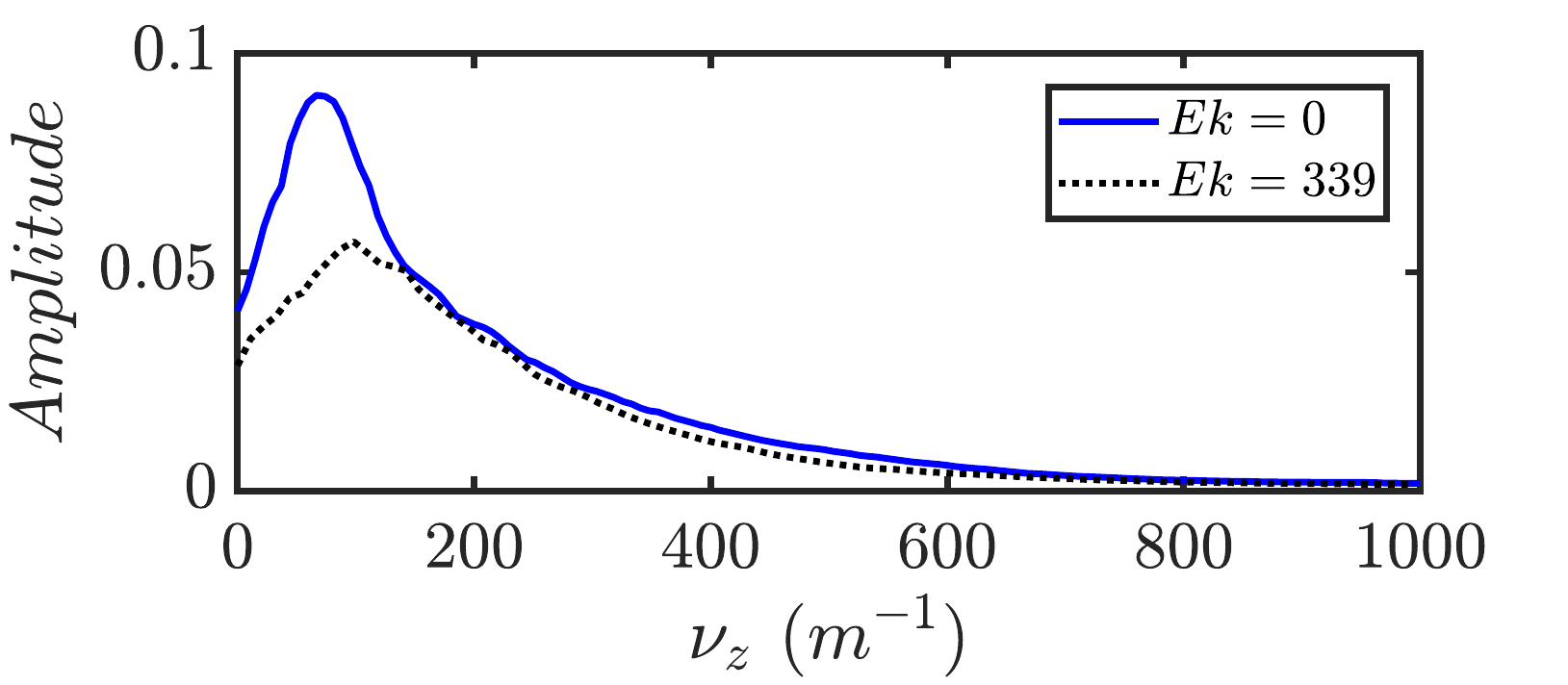}
	\label{fig:fig_5e}
	\end{subfigure}
	\begin{subfigure}{0.49\linewidth}
	\centering
	\caption{\hspace*{-1.5em} \vspace{-0.5em}}
	\includegraphics[width = \textwidth]{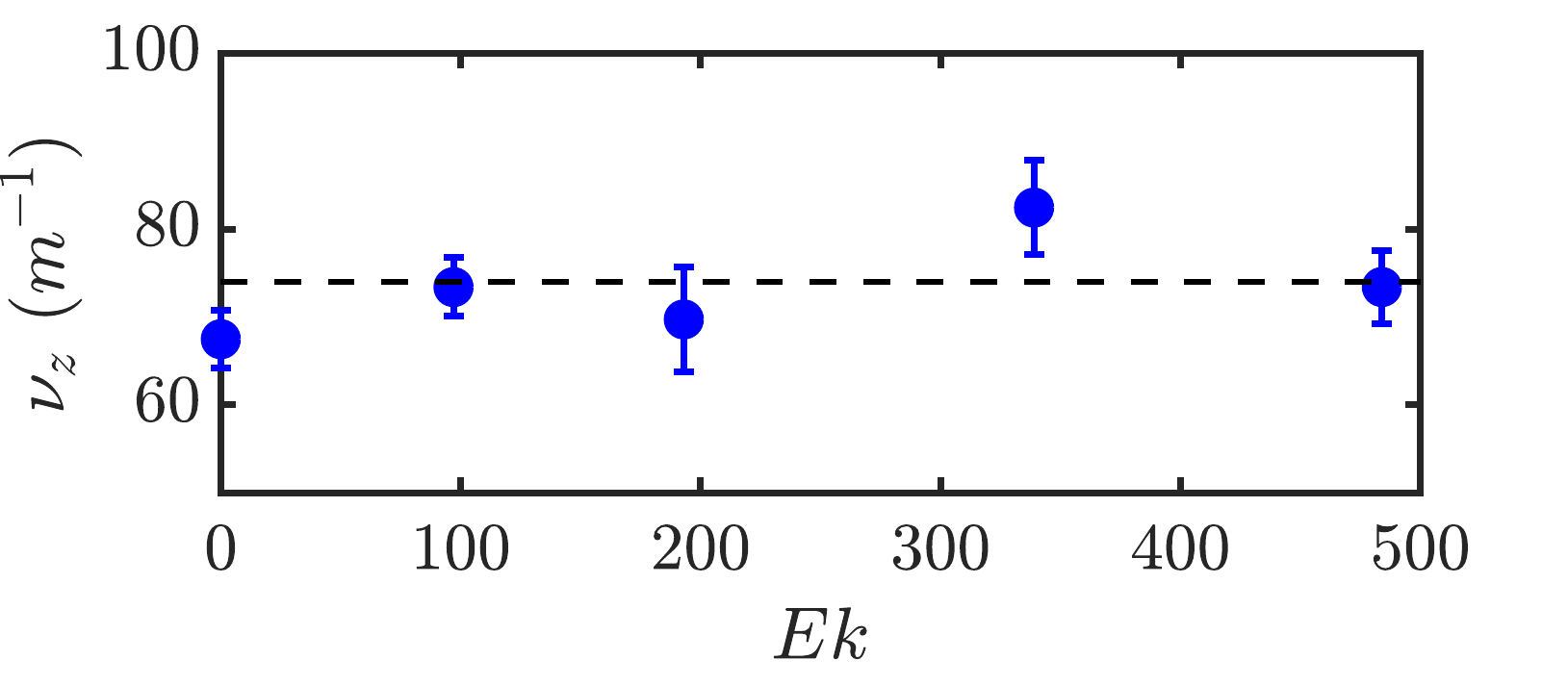}
	\label{fig:fig_5f}
	\end{subfigure}
	\caption{Single case $(\tilde{t} = 955)$ input for FFT for a) $Ek = 0$, b) $Ek = 339$; single case FFT output result for c) $Ek = 0$, d) $Ek = 339$; e) time-averaged FFT distributions for $Ek = 0$ and $Ek = 339$; f) time-averaged axial wavenumber at each $Ek$. The rest of the parameters remain unchanged from Fig. 2.}
	\label{fig:fig5}
\end{figure}

We have also performed fast Fourier transforms (FFT) of the time-averaged interface profiles in the wavy regime (for $z>L_{2D}$ for every $Ek$ value considered) after subtracting the mean film height; this is shown in Fig. \ref{fig:fig5}. This subtraction is to account for the $\nu_z = 0 ~\text{m}^{-1}$ wavenumber that would be prevalent with a signal with a non-zero mean amplitude. The time-averaging was performed for times that exceeded those associated with the establishment of a dynamic steady-state for each $Ek$. %
In Figs. \ref{fig:fig_5a} and \ref{fig:fig_5b}, we show example snapshots of the spatial development of the interface for $Ek=0$ and $Ek=339$, with their power spectra shown in Figs. \ref{fig:fig_5c} and \ref{fig:fig_5d}, respectively. 
It is clearly seen that the profile associated with $Ek=0$ are significantly wavier than its higher $Ek$ counterpart, characterised by large-amplitude peaks preceded by high-wavenumber capillary waves. 
This is reflected by the higher energy content of the wavenumber modes, particularly at relatively low $\nu_z$, as demonstrated via comparison of Figs. \ref{fig:fig_5c} and \ref{fig:fig_5d}. 
Applying a moving average allows us to compare the power spectra associated with the two $Ek$ values  %
in Fig. \ref{fig:fig_5e}. This shows that the dominant mode is weakly-dependent on $Ek$ though the associated amplitude is larger in the purely falling film case. 
In Fig. \ref{fig:fig_5f}, we plot the variation of the wavenumber associated with the dominant mode with $Ek$. %
It is seen that the dominant modes have wavenumbers that %
are in the range 70 - 85 m$^{-1}$ for the range of $Ek$ studied, with no strong dependence on $Ek$ suggesting that the structure in the axial direction is not dominated by the centrifugal forces due to cylinder rotation.  %

\begin{figure}[htbp!]
	\begin{subfigure}{0.67\linewidth}
	\centering
	\caption{\hspace*{-1.5em} \vspace{-0.5em}}
	\includegraphics[width = \linewidth]{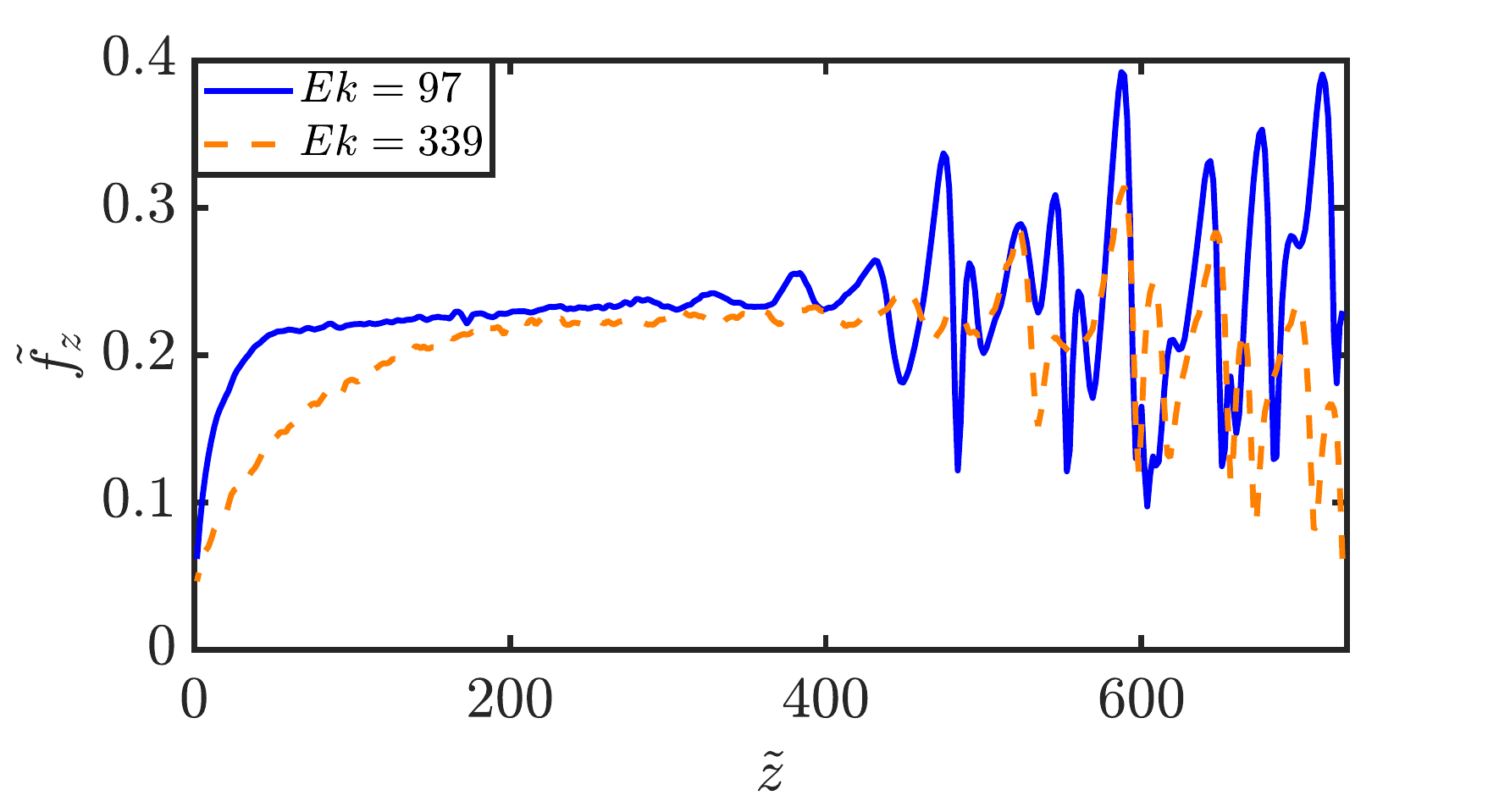}
	\label{fig:fig_6a}
	\end{subfigure}
	\begin{subfigure}{0.67\linewidth}
	\centering
	\caption{\hspace*{-1.5em} \vspace{-0.5em}}
	\includegraphics[width=\linewidth]{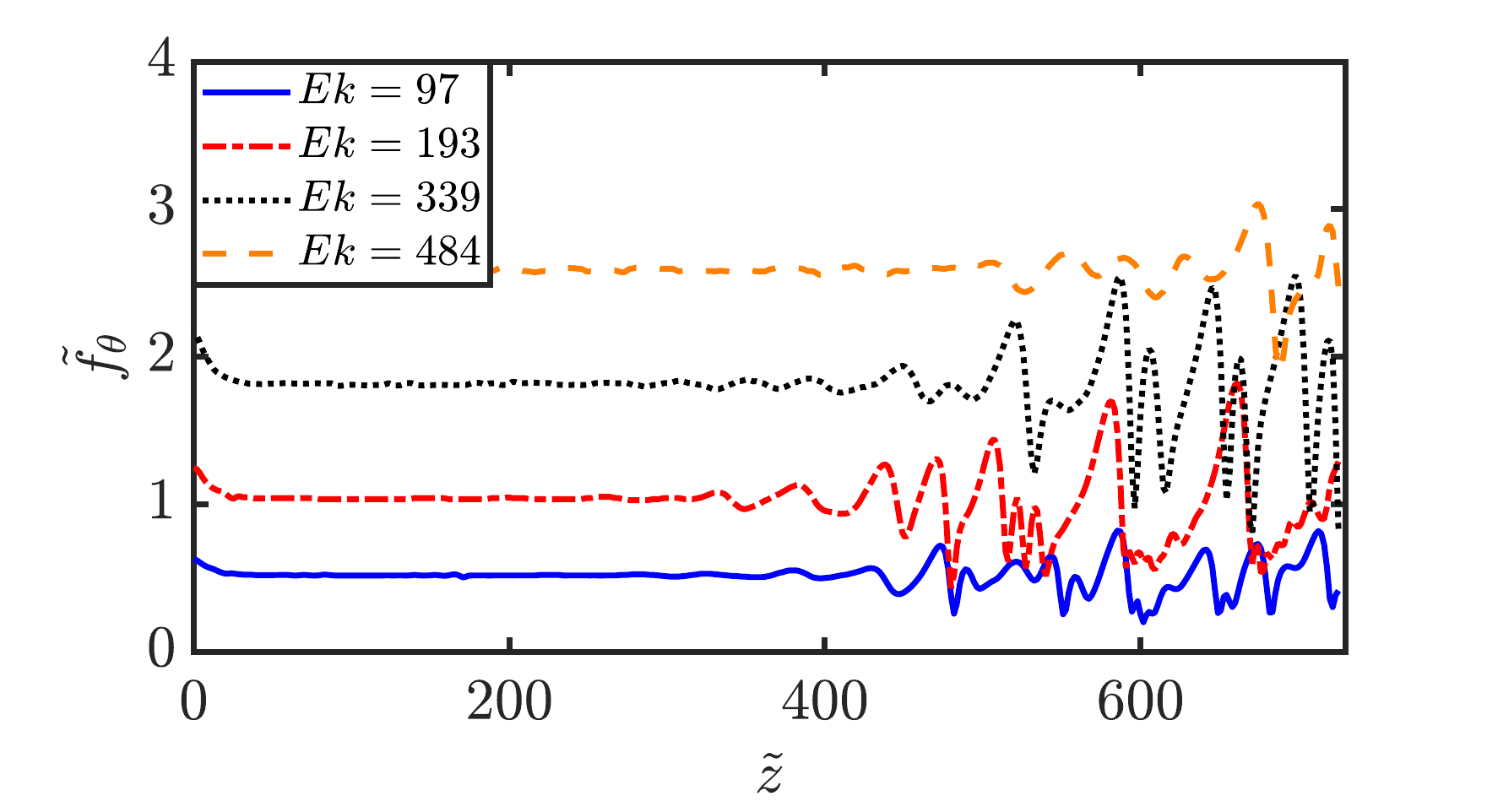}
	\label{fig:fig_6b}
	\end{subfigure}
	\caption{Axial (a) and azimuthal (b) flow rates for a range of $Ek$ at $\tilde{t} = 1193$. The rest of the parameters remain unaltered from Fig. 2. }
	\label{fig:fig6}
\end{figure}

We analyse the flow further in both the $z$- and $\theta$-directions by plotting the axial and 
azimuthal flow rates, which are respectively expressed by 
\begin{equation}
\tilde{f}_z      = \int^{1/\beta}_{1/\beta - \tilde{h}} \tilde{r}\tilde{u}_z d\tilde{r}, ~~~~~
\tilde{f}_\theta = \int^{1/\beta}_{1/\beta - \tilde{h}} \tilde{r}\tilde{u}_\theta d\tilde{r},
\end{equation}
where $\tilde{u}_z$ and $\tilde{u}_\theta$ denote the axial and azimuthal velocity components, respectively, and
$(\tilde{f}_z,\tilde{f}_\theta)=(f_z,f_\theta)/2\pi h^2_N u_N$. 
In Fig. \ref{fig:fig_6a}, we plot the axial variation of $\tilde{f}_z$ for the falling film and $Ek=339$ cases where it is seen that the waviness in $\tilde{f}_z$, which closely relates to that of the interface, is suppressed for the high $Ek$ case; here, the rest of the parameters remain unchanged from Fig. \ref{fig:fig2}. This result is matched by that depicted in Fig. \ref{fig:fig_6b} in which it is shown that 
the average $\tilde{f}_\theta$ increases with $Ek$ due to the rise in the influence of centrifugal forces on the flow. Despite this trend, close inspection of Fig. \ref{fig:fig_6b} also reveals that the amplitude of the oscillations in $\tilde{f}_\theta$ is maximised for an intermediate range of $Ek$. 
This is because %
an increase in $Ek$ results in an increase in $\tilde{u}_\theta$, which promotes $\tilde{f}_\theta$, but also leads to suppression of interfacial waviness. These competing effects give rise to the results presented in Fig. \ref{fig:fig_6b}. %

\begin{figure}
    \begin{subfigure}[b]{0.49\linewidth}
    \centering
    \caption{\hspace*{-3em} \vspace{-0.2em}}
    \includegraphics[width=0.9\textwidth]{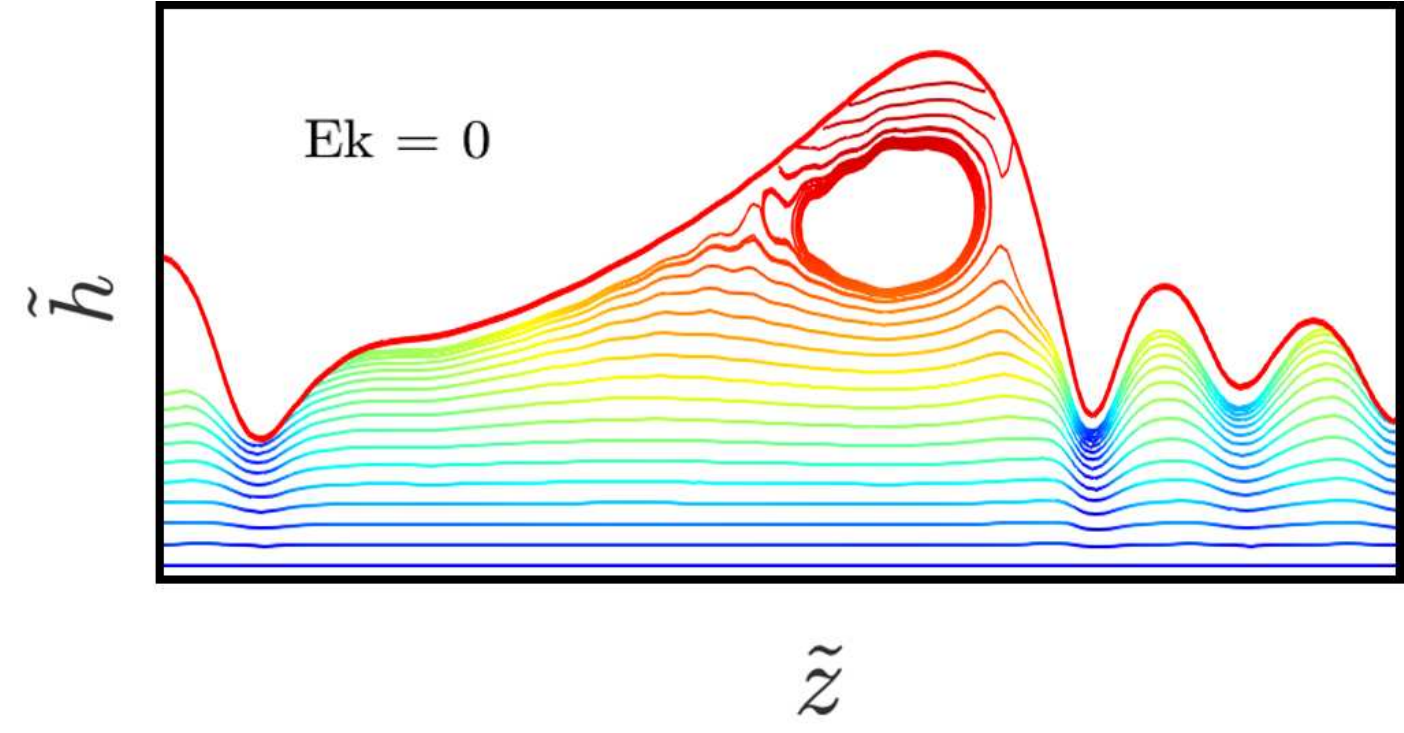}
    \label{fig:fig6A_1}
    \end{subfigure}
    \begin{subfigure}[b]{0.49\linewidth}
    \centering
    \caption{\hspace*{-3em} \vspace{-0.2em}}
    \includegraphics[width=0.95\textwidth]{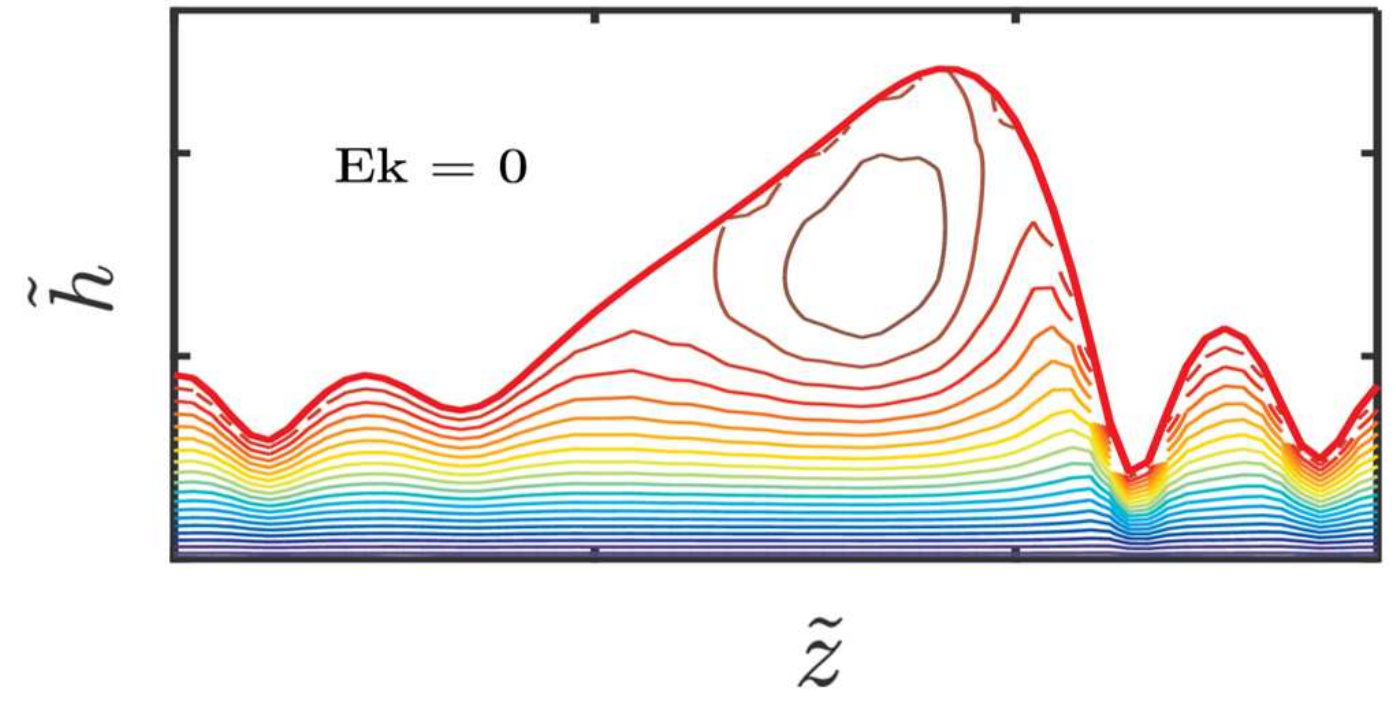}
    \label{fig:fig6B_1}
    \end{subfigure}
    \begin{subfigure}[b]{0.49\linewidth}
    \centering
    \caption{\hspace*{-3em} \vspace{-0.2em}}
    \includegraphics[width=0.9\textwidth]{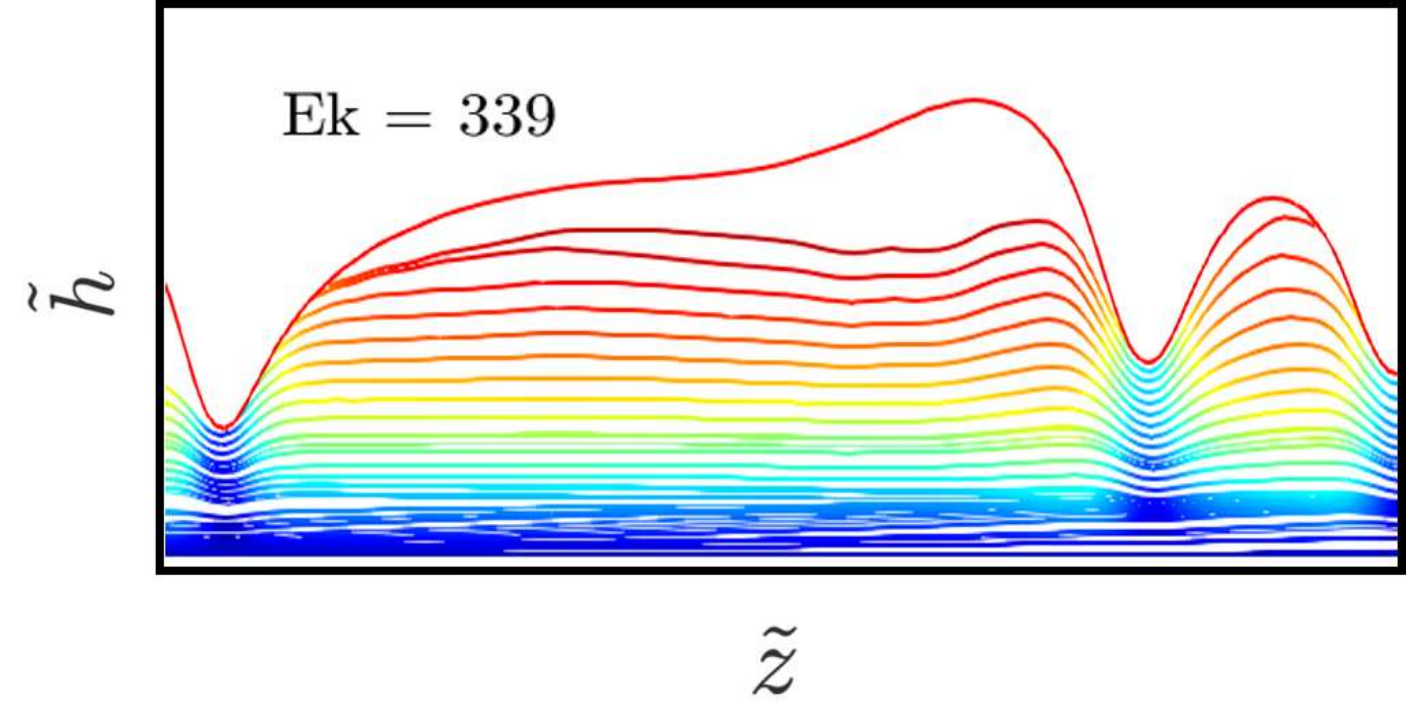}
    \label{fig:fig6C_1}
    \end{subfigure}
    \begin{subfigure}[b]{0.49\linewidth}
    \centering
    \caption{\hspace*{-3em} \vspace{-0.2em}}
    \includegraphics[width=0.92\textwidth]{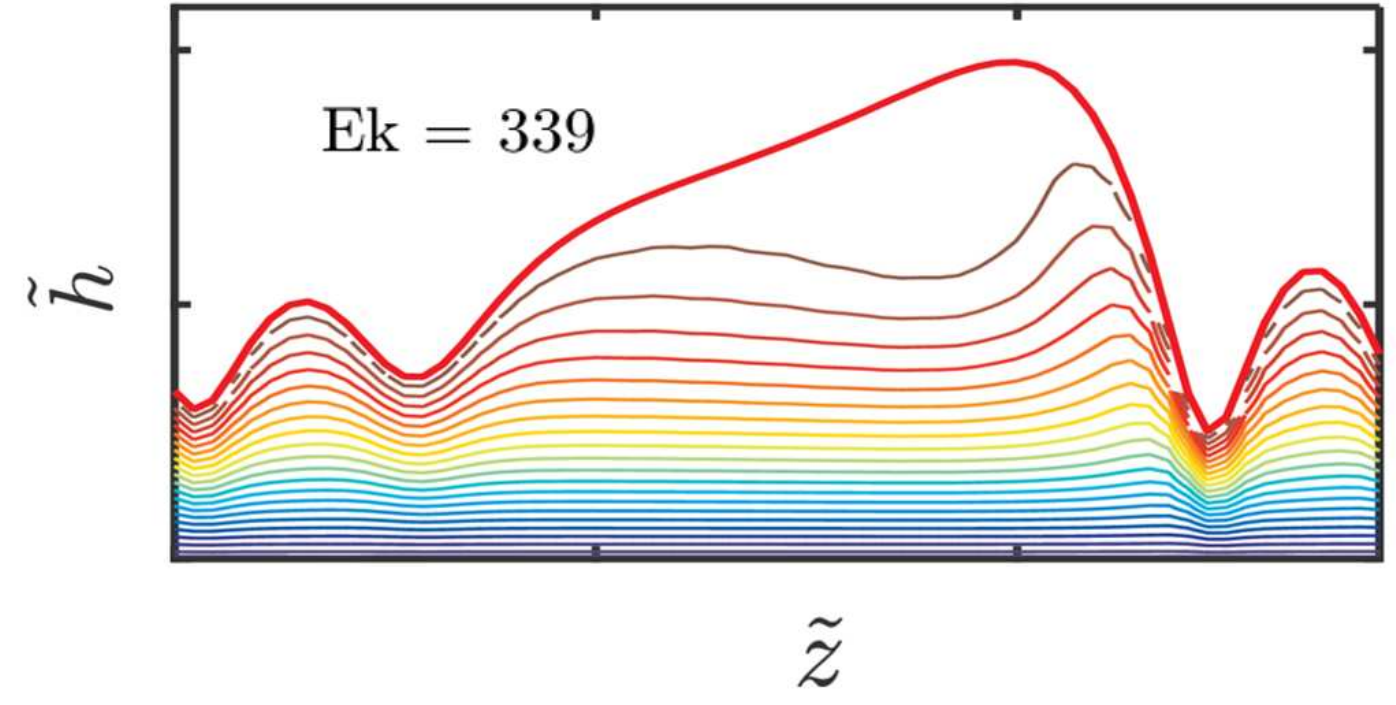}
    \label{fig:fig6D_1}
    \end{subfigure}
    \begin{subfigure}[b]{0.49\linewidth}
    \centering
    \caption{\hspace*{-1.5em} \vspace{-0.7em}}
    \includegraphics[width=\textwidth]{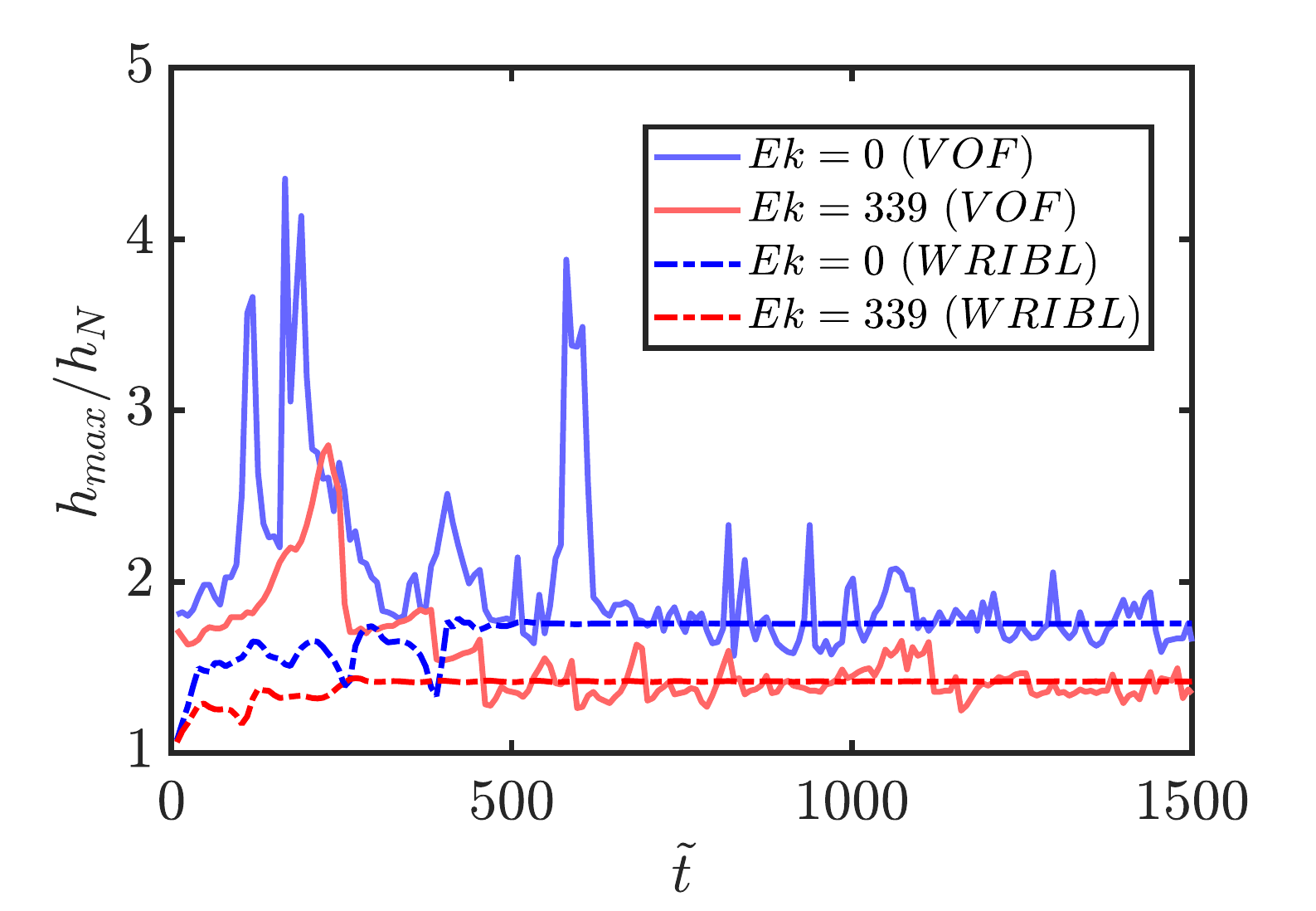}
    \label{fig:fig6E_1}
    \end{subfigure}
    \begin{subfigure}[b]{0.49\linewidth}
    \centering
    \caption{\hspace*{-1.5em} \vspace{-0.7em}}
    \includegraphics[width=0.95\textwidth]{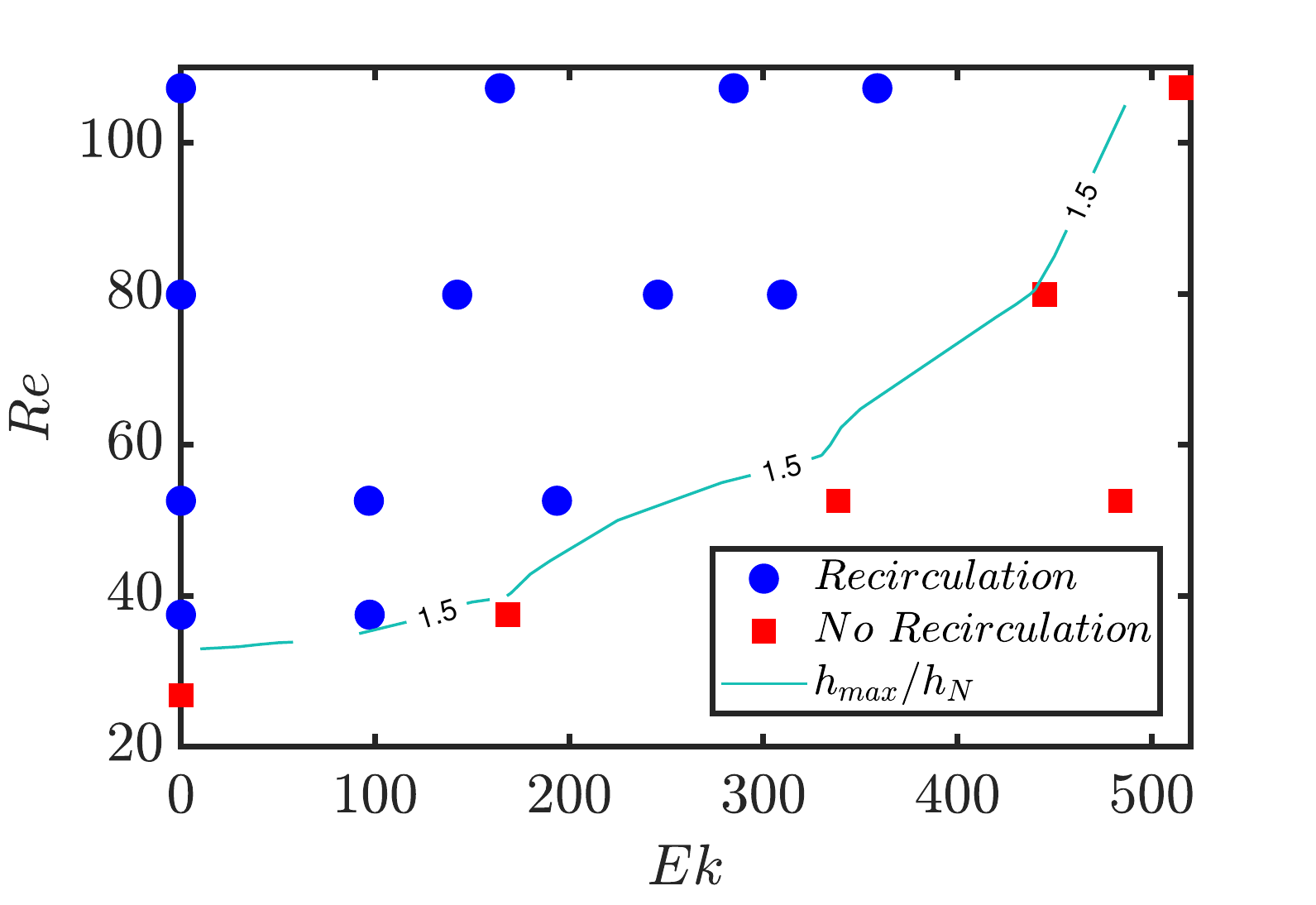}
    \vspace{1.6em}
    \label{fig:fig6F_1}
    \end{subfigure}
    \caption{Comparison of 2D wave formation for a) VOF, $Ek = 0$; b) WRIBL, $Ek = 0$; c) VOF, $Ek =339$; and d) WRIBL, $Ek = 339$; e) temporal evolution of the maximum film height normalised by the Nusselt film thickness, $h_{max}/h_N$, generated via the VOF and WRIBL methods; f) $Re - Ek$ phase diagram showing the boundary separating waves wherein recirculation is observed in a frame-of-reference moving with the wave celerity and for which $h_{max}/h_N \approx 1.5$.  The rest of the parameters remain unaltered from Fig. 2.}
    \label{fig:fig6_1}
\end{figure}

The presence of recirculation is an important feature of falling films, with relevance in industrial applications due to  enhanced mixing. We have shown that the presence of rotation has a marked effect on the shape of the waves and hence the degree and presence of recirculation. The interplay between the inertial instability in the axial direction and the centrifugal stabilisation determines the wave formation. It is instructive to construct a phase diagram with these parameters determining whether recirculation is present, for which we require a model to optimise processing time, compared to the use of full DNS. Rohlfs {\it{et al.}} \cite{rohlfs_swx_2018} have recently developed %
{\it{WaveMaker}}, a Matlab-based software, which simulates periodic waves in 2D and 3D domains, utilising a weighted residual integral boundary layer (WRIBL) approach. Both domains are utilised and compared to the results from the VOF DNS approach, with the 2D full-second order WRIBL model used to construct a phase diagram in  $Re - Ek$ space.

One notes that the VOF simulations are of a developing wave and are periodic in the azimuthal direction, compared to the WRIBL model, which is for spatially-periodic waves in both the streamwise and spanwise directions. However, a comparison can be made between more developed waves in the latter part of the domain $\tilde{z} > L_{2D}$ and those created via the  WRIBL approach. A domain size of 0.0143 m was chosen based on an axial wavenumber that has been shown to be $\nu_z \approx 70~ \text{m}^{-1}$ for the range of $Ek$ investigated in Fig. \ref{fig:fig_5f}. We employ the analogy that the simulation in the case of a film on the inside of a rotating cylinder corresponds to that of its counterpart on a plane inclined at an angle $\gamma$ set by $\gamma = \tan^{-1}\left(g/\Omega^2R\right)$.
For a chosen $Ek$ value, the corresponding $\Omega$ is then used to determine $\gamma$ whence the Reynolds number for the inclined plane WRIBL simulation is given by
\begin{equation}
    Re_{\rm{WRIBL}}=\frac{h_{N}^{3}}{3 v^{2}}g\sin \gamma. %
\end{equation}
It is important to check the curvature of the cylinder in the VOF simulations is such that an inclined plane analogy is valid. Chen {\it{et al.}} \cite{chen_ijhmt_2004} found that the influence of the curvature on stability was negligible for $1/\beta >> 10$; in the present case, $1/\beta = 146$, which demonstrates the validity of our approach.

Figs. \ref{fig:fig6A_1} and Fig. \ref{fig:fig6B_1} provide a comparison for $Ek = 0$ of the 2D waves from the VOF and WRIBL simulations, respectively. With no centrifugal force suppressing wave formation, the recirculation within the solitary wave is evident. Furthermore, the capillary waves are significantly smaller than the preceding main hump. In contrast, at $Ek = 339$, as observed in Fig. \ref{fig:fig6C_1} and Fig. \ref{fig:fig6D_1}, recirculation is absent from the main wave hump. %
We then compare the WRIBL results to those obtained from the VOF simulations in terms of temporal evolution of the maximal film thickness normalised by the Nusselt film thickness, $h_{max}/h_N$. 
Fig. \ref{fig:fig6E_1} shows that the WRIBL results converge to a steady-state, with a lower $h_{max}/h_N$ for the higher $Ek$ case, due to the increased stability. This $h_{max}/h_N$ for each $Ek$ matches well with the quasi-steady state obtained from the VOF simulations, suggesting that the developed waves where $\tilde{z} >> L_{2D}$ are comparable to those obtained by the WRIBL model, further supported qualitatively through Figs. \ref{fig:fig6A_1} - \ref{fig:fig6D_1}. 
These results indicate that the WRIBL approach provides a reasonably good approximation of the VOF DNS predictions. On this basis, the WRIBL method was used to construct the phase diagram shown in 
Fig. \ref{fig:fig6F_1}. %
Notably, the level of inertia, characterised by $Re$, required to have recirculation increases with $Ek$. The critical boundary for recirculation in $Re-Ek$ space coincides with the %
$h_{max}/h_N = 1.5$ contour; thus, the latter  provides an effective 
criterion for the onset of recirculation. As will be discussed below, this can be applied in the three-dimensional case to determine the presence of recirculation, to which we now turn our attention.

\begin{figure}[!htbp]
	\begin{minipage}{0.24\textwidth}
	\begin{subfigure}[b]{\linewidth}
		\centering
		\caption{}
		\includegraphics[width = \textwidth]{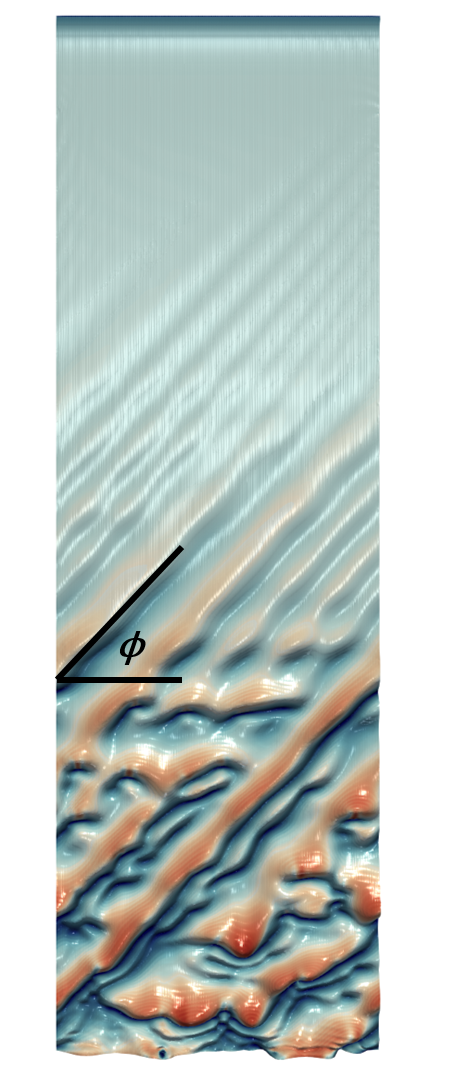}
		\label{fig:fig_7a}
	\end{subfigure}
	\end{minipage}
	\begin{minipage}{0.205\textwidth}
	\begin{subfigure}[b]{\linewidth}
		\centering
		\vspace{0.5em}
		\caption{\vspace{-0.6em}}
		\includegraphics[width = \textwidth]{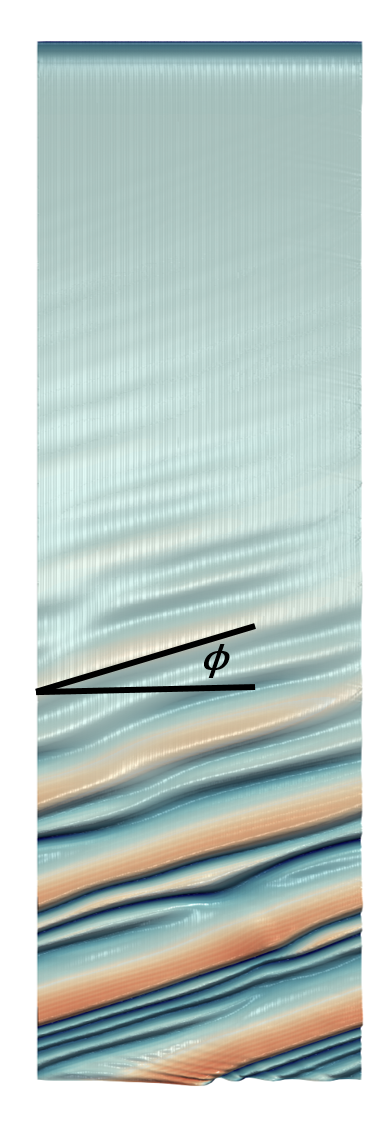}
		\label{fig:fig_7b}
	\end{subfigure}
	\end{minipage}
	\begin{minipage}{0.49\textwidth}
		\vspace{2em}
	\begin{subfigure}[b]{\linewidth}
		\centering
		\caption{\hspace{-2em}\vspace{-0.5em}}
		\includegraphics[width = \textwidth]{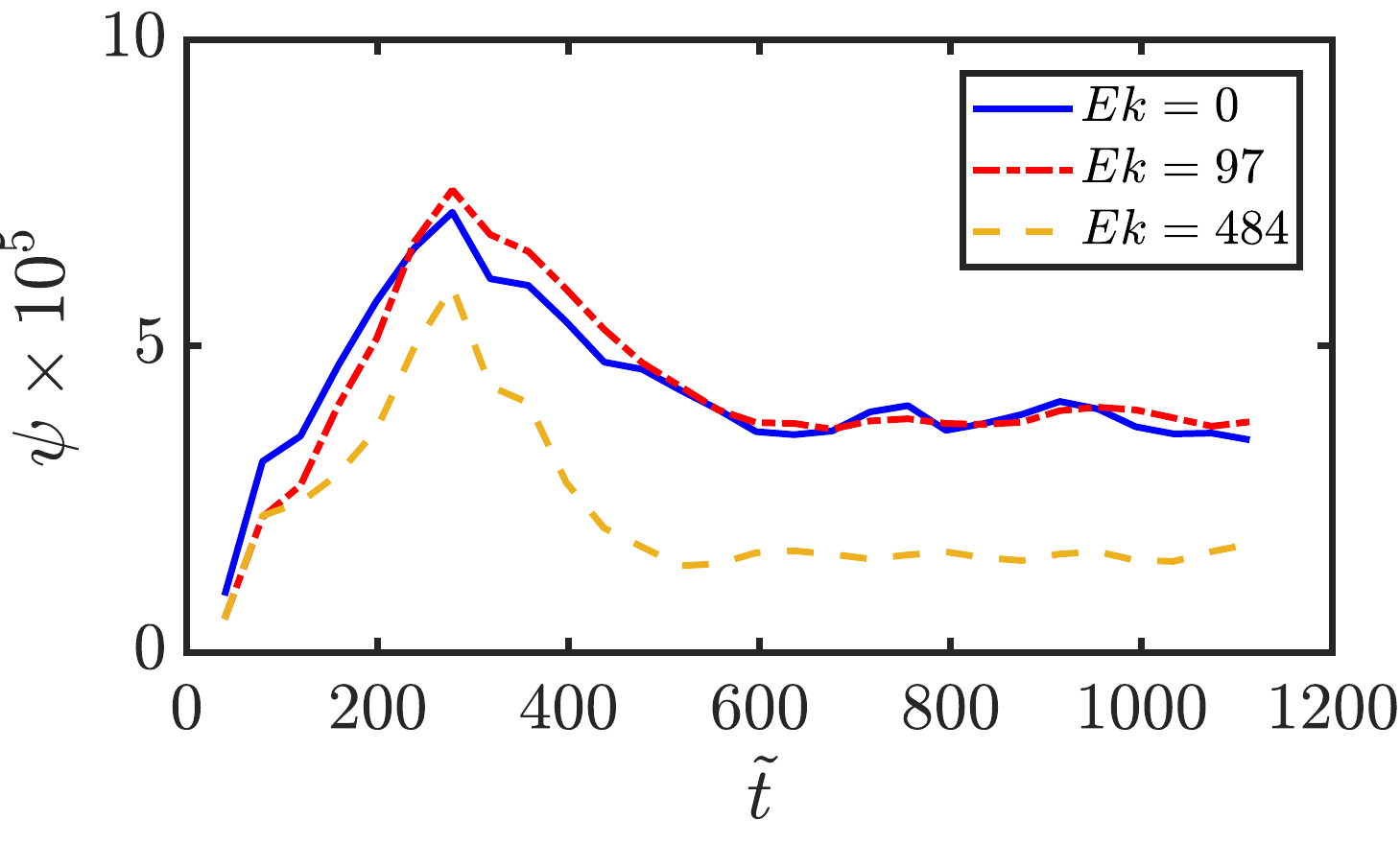}
		\label{fig:fig_7c}
	\end{subfigure}
	\begin{subfigure}[b]{\linewidth}
		\centering
		\caption{\hspace{-2em}\vspace{-0.5em}}
		\includegraphics[width = \textwidth]{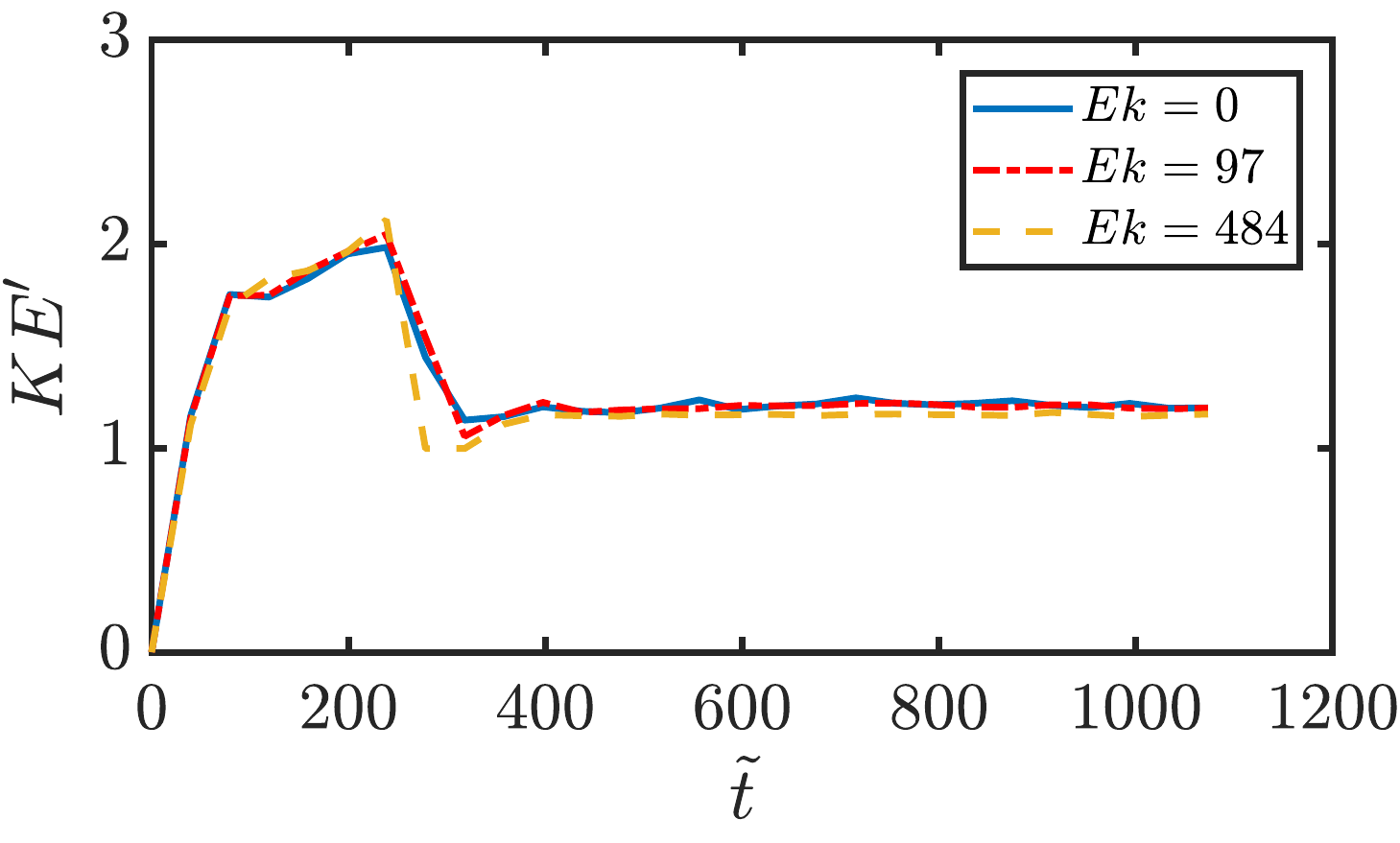}
		\label{fig:fig_7d}
	\end{subfigure}
	\end{minipage}
	\begin{subfigure}[b]{0.49\linewidth}
		\centering
		\caption{}
		\hspace{-2em}
		\vspace{1em}
		\includegraphics[width = \textwidth]{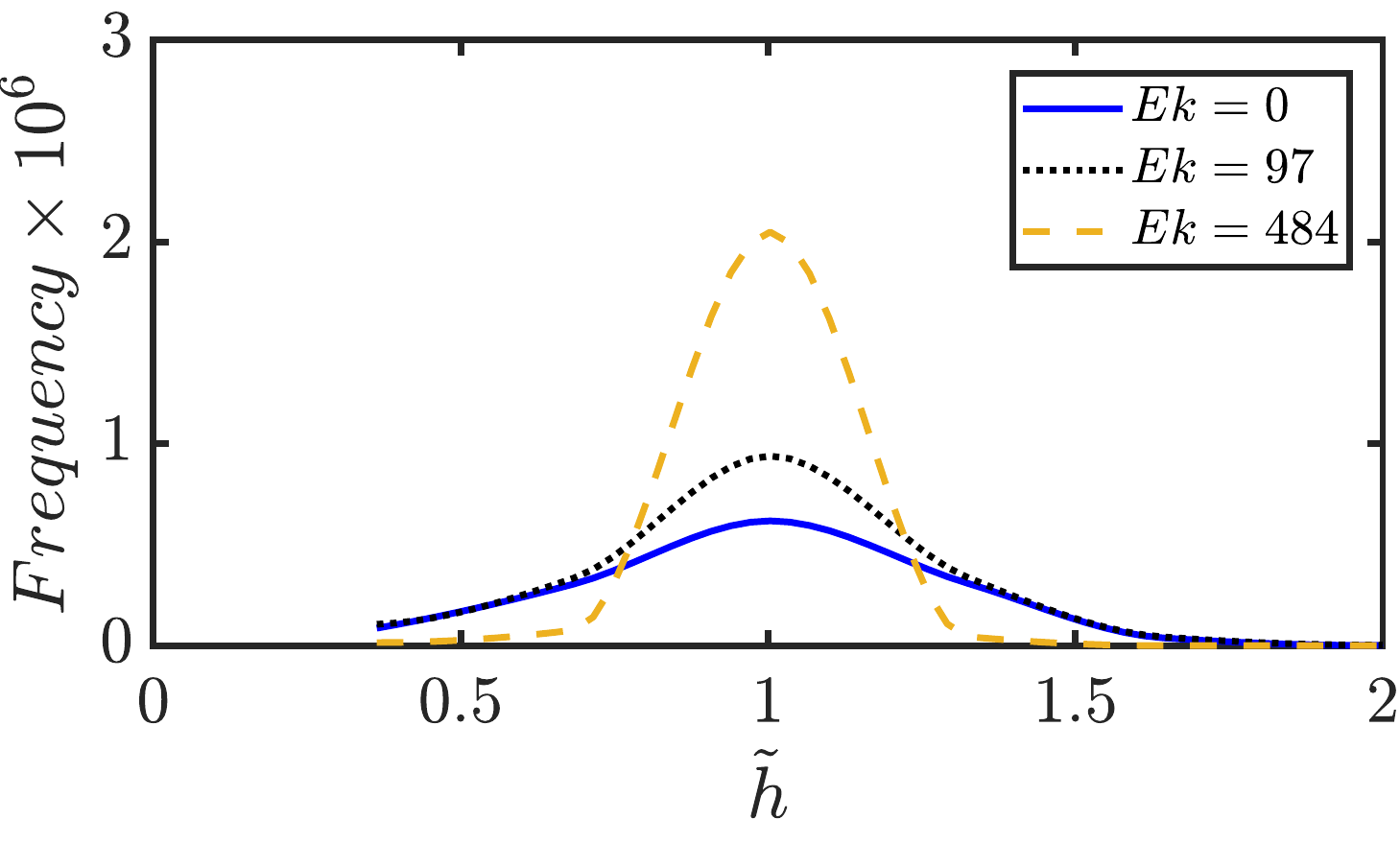}
		\label{fig:fig_7e}
	\end{subfigure}
	\begin{subfigure}[b]{0.49\linewidth}
		\centering
		\caption{\hspace*{-1.5em}}
		\hspace{-2em}
		\vspace{1em}
		\includegraphics[width = \textwidth]{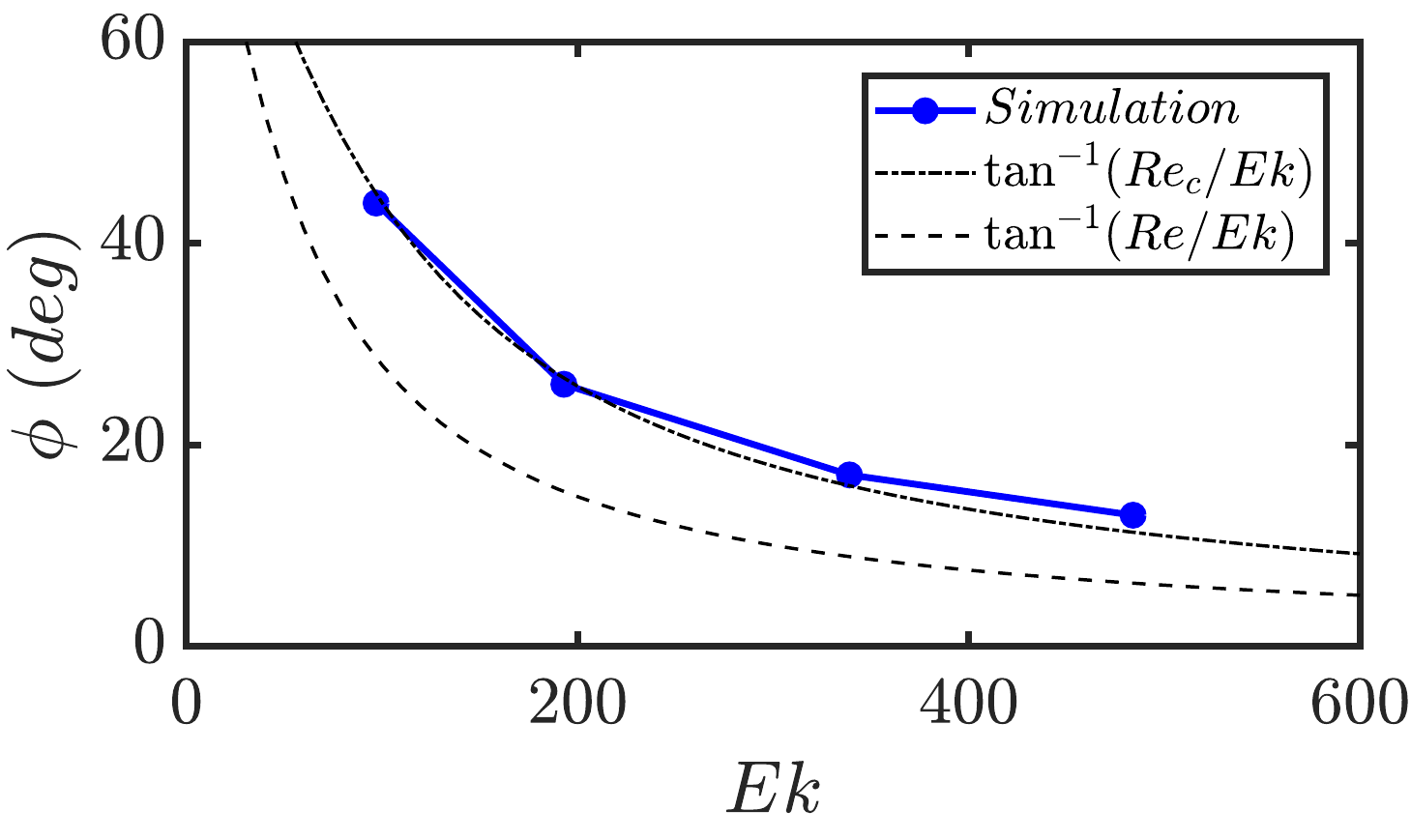}
		\label{fig:fig_7f}
	\end{subfigure}
	\caption{Snapshots taken at $\tilde{t} = 955$ showing contour plots of the interface that demonstrate the formation of angled waves for a) $Ek = 97$ and b) $Ek = 339$; dynamic evolution of the film waviness $\psi$ and kinetic energy $KE'$, shown in (c) and (d), respectively, for a range of $Ek$; e) narrowing of the film height distribution with an increase in $Ek$. The rest of the parameters remain unaltered from Fig. 2; %
	f) comparison of the angle $\phi$ values obtained from simulations with those determined from Eq. (\ref{eq:phi_a}) with $\cal U$ from extracted wave celerities, and Nusselt theory.}
	\label{fig:fig7}
\end{figure}

\section{Three-dimensional Simulations}\label{section_4}
In this section, we present the results of our 3D simulations of the interfacial dynamics. Our aim is to uncover details of the flow masked by the axisymmetric assumption made in the previous section; 
details of the simulation setup are found in \ref{section_2}.  %
In Figs. \ref{fig:fig_7a} and \ref{fig:fig_7b}, we show snapshots of the interface in the `unwrapped' $(\theta,z)$ plane for $Ek=97$ and $Ek=339$, respectively, wherein the colour represents the thickness of the film. Similarly to the 2D predictions discussed in Section III, these figures show clearly the development of a wavy interface from an essentially waveless region near the domain inlet. In 3D, however, it is seen that  
the effect of rotation causes the formation of waves that appear to be oriented to the horizontal with a reasonably well-defined angle $\phi$. In Fig. \ref{fig:fig_7c}, it is seen that the film waviness, $\psi$, decreases with $Ek$, just as in the 2D case, and its temporal variation reaches a steady-state beyond a dimensionless time whose value is not a strong function of $Ek$. In Fig. \ref{fig:fig_7d}, we also see that the kinetic energy, which, just as in the 2D case, is weakly-dependent on $Ek$, also reaches a steady-state albeit at an earlier time than $\psi$.
The distribution of film heights is also obtained from the 3D simulations and shown in Fig. \ref{fig:fig_7e}. We observe a narrowing of the distribution around the Nusselt film height as $Ek$ increases due to the stabilising effect of an increase in the rotational speed, as the suppression of wave formation reduces fluctuations around $h_N$.

\begin{figure}[!htbp]
	\centering
	\begin{minipage}{0.4\textwidth}
	\begin{subfigure}[b]{\linewidth}
		\centering
		\caption{\vspace{-0.5em}}
		\includegraphics[width = \textwidth]{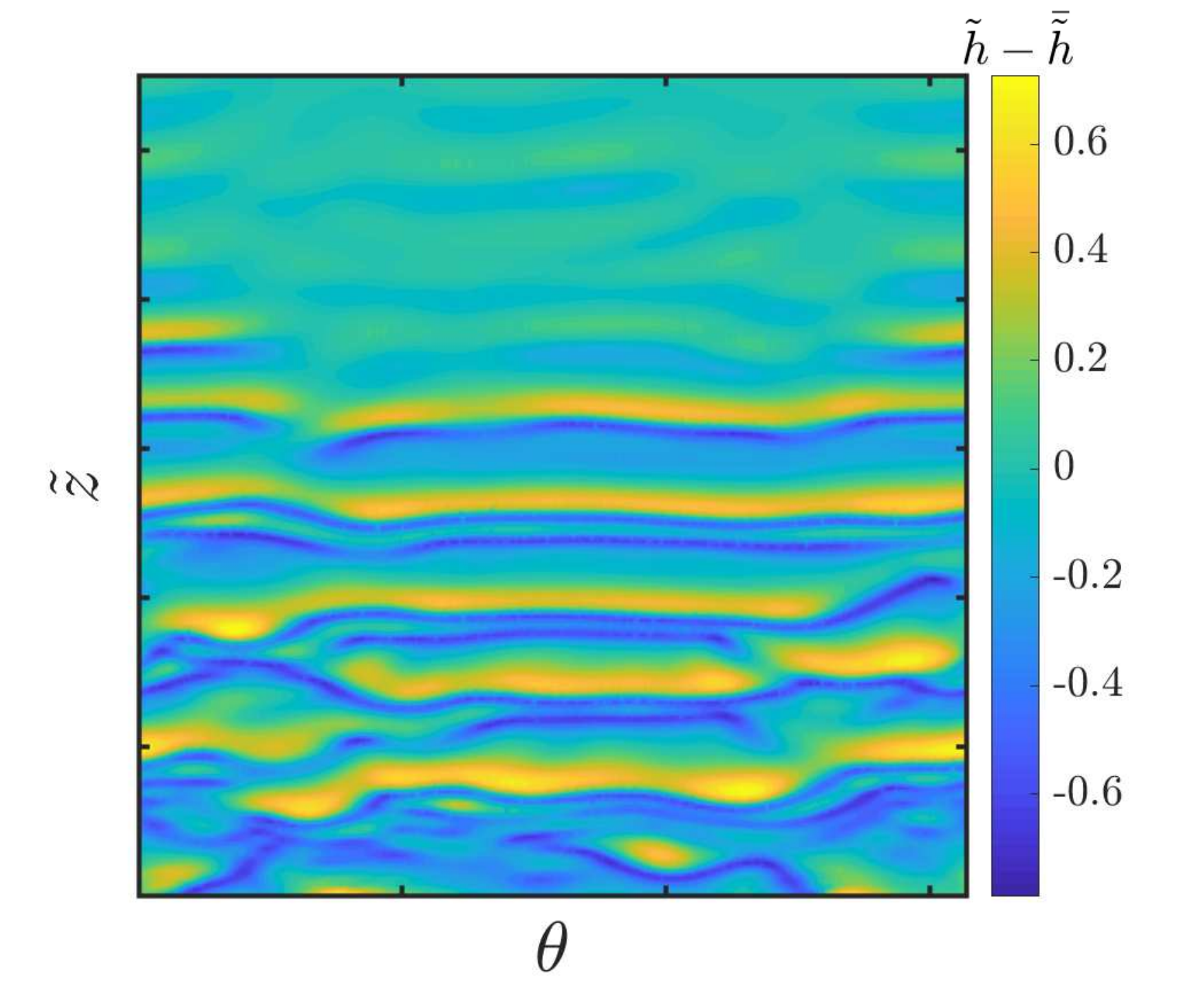}
		\label{fig:fig_8a}
	\end{subfigure}	
	\end{minipage}
	\begin{minipage}{0.4\textwidth}
	\begin{subfigure}[b]{\linewidth}
		\centering
		\caption{\hspace*{0.5em} \vspace{-0.5em}}
		\includegraphics[width = \textwidth]{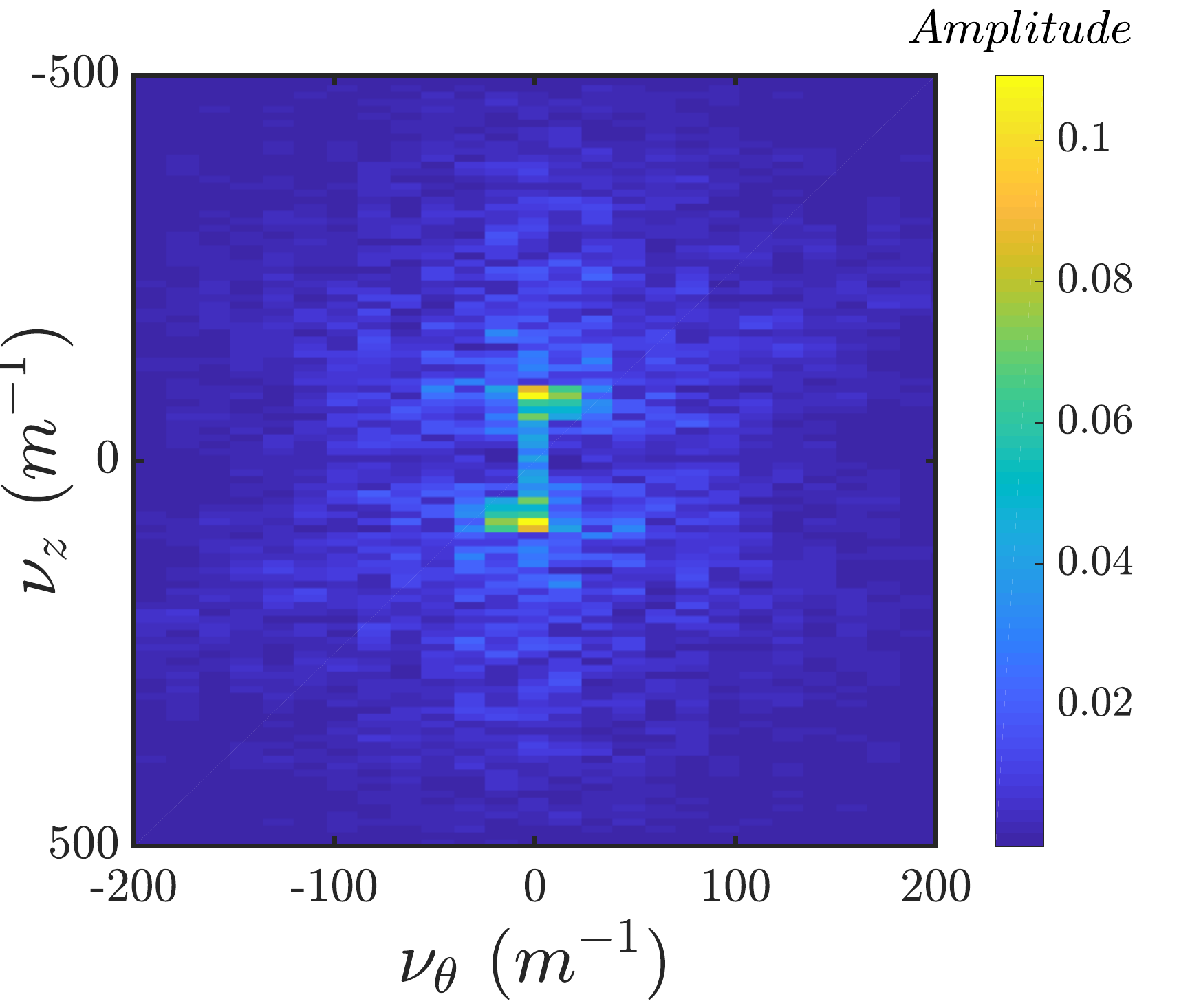}
		\label{fig:fig_8b}
	\end{subfigure}	
	\end{minipage}
	\begin{minipage}{0.4\textwidth}
	\begin{subfigure}[b]{\linewidth}
		\centering
		\caption{\vspace{-0.5em}}
		\includegraphics[width = \textwidth]{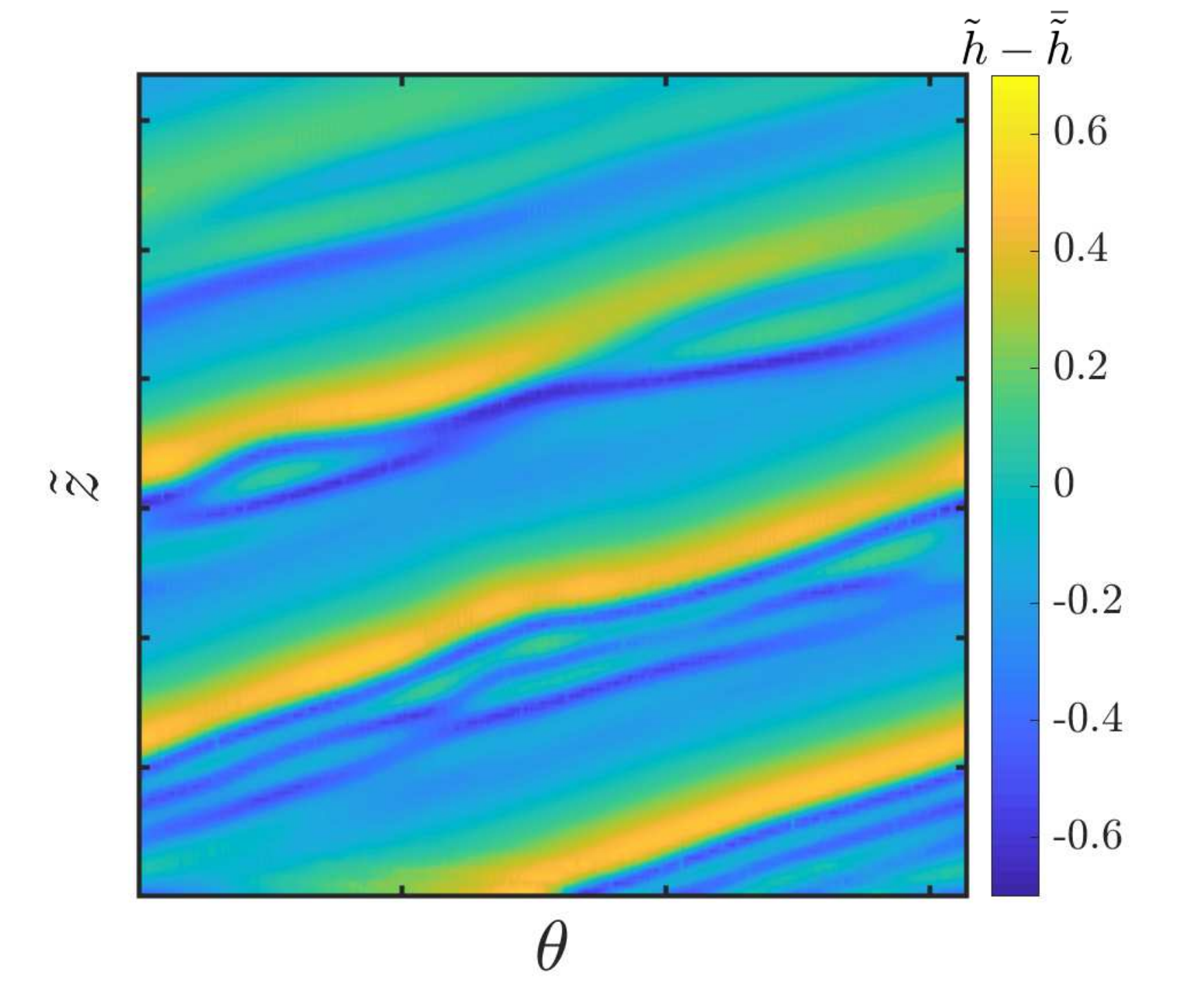}
		\label{fig:fig_8c}
	\end{subfigure}
	\end{minipage}
	\begin{minipage}{0.4\textwidth}
	\begin{subfigure}[b]{\linewidth}
			\caption{ \hspace*{0.5em} \vspace{-0.5em}}
			\includegraphics[width = \textwidth]{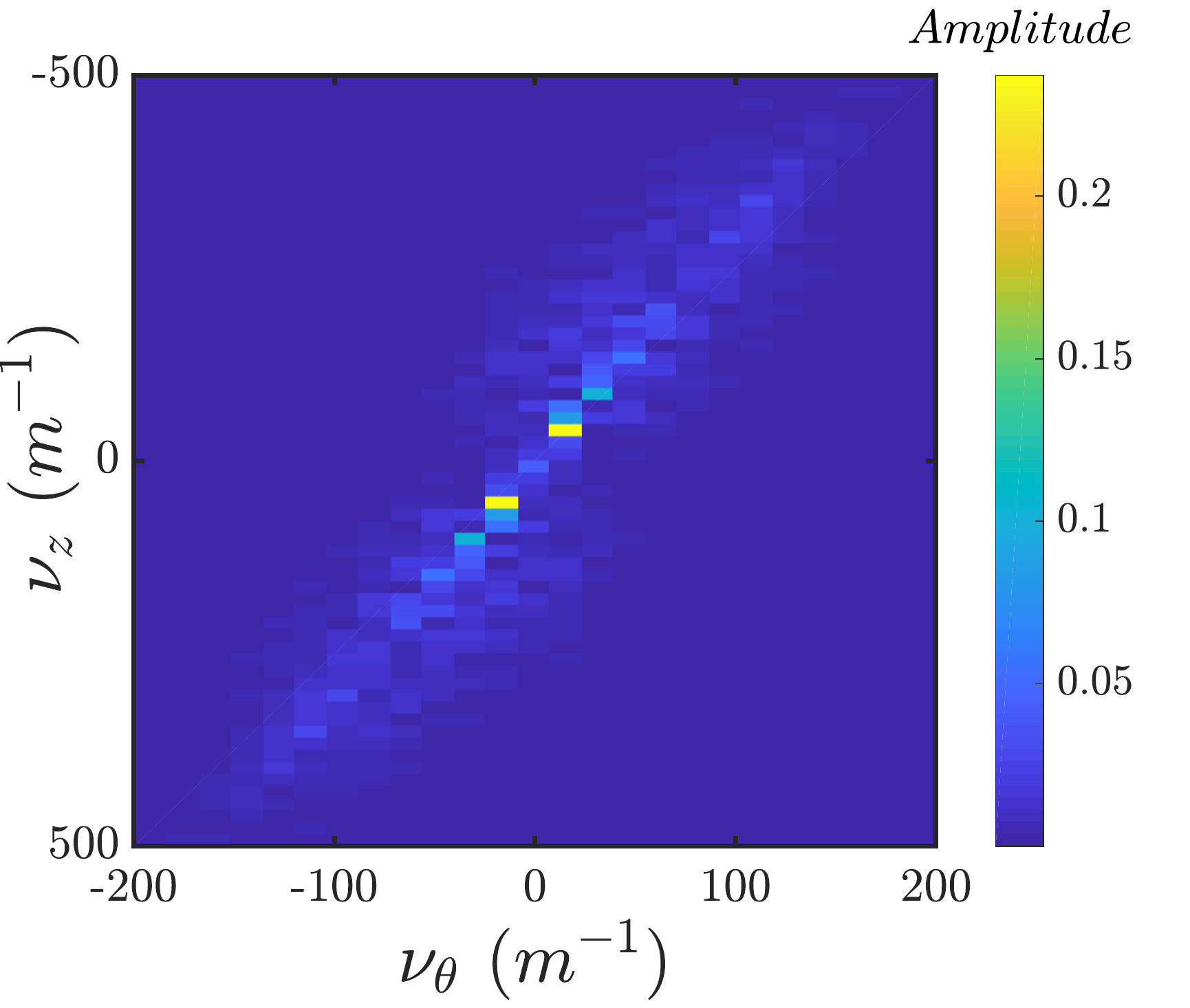}
			\label{fig:fig_8d}
		\end{subfigure}
	\end{minipage}
	\begin{minipage}{0.6\textwidth}
		\begin{subfigure}[b]{\linewidth}
			\centering
			\caption{\hspace*{-1.8em} \vspace{-0.5em}}
			\includegraphics[width = \textwidth]{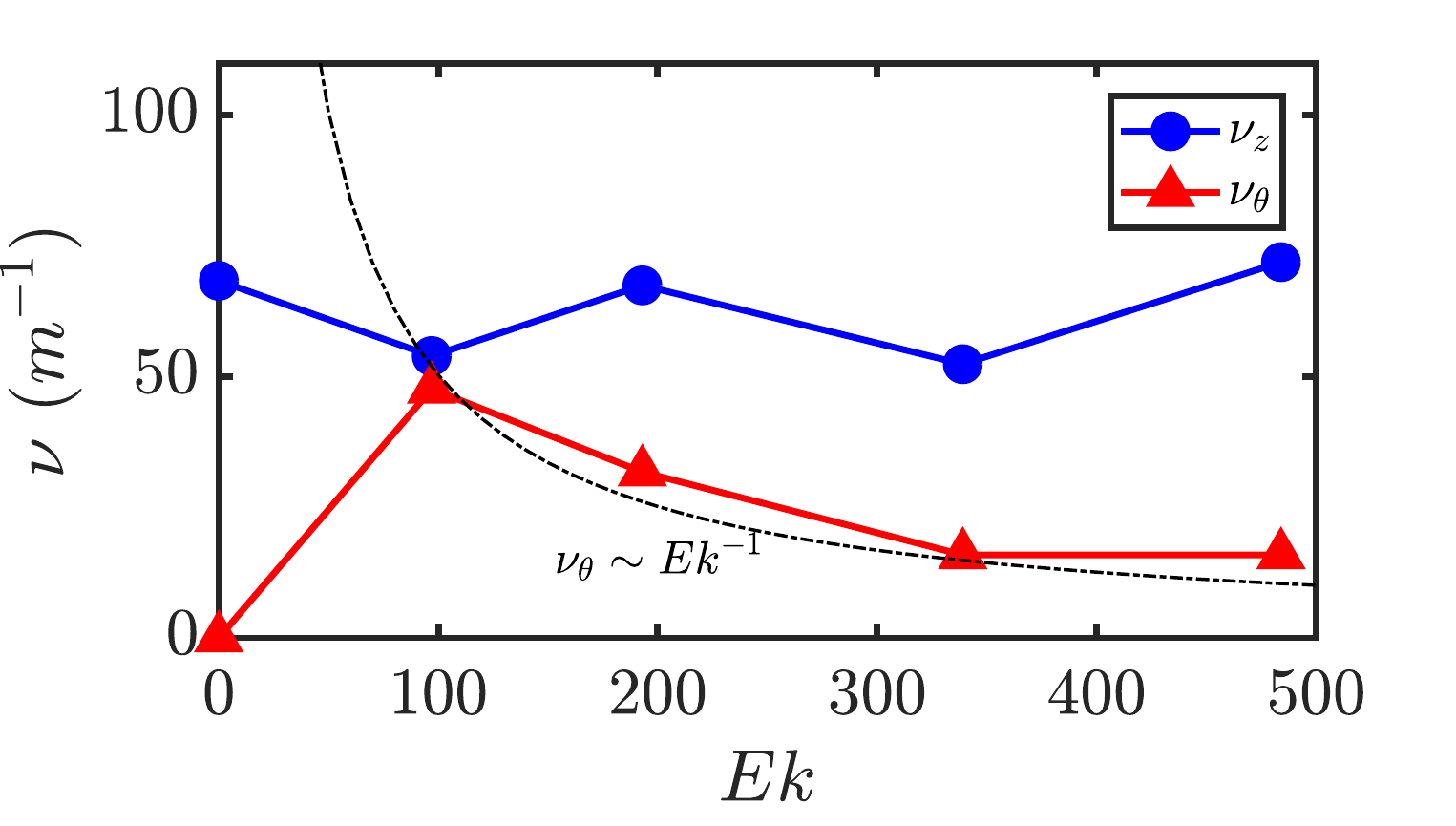}
			\label{fig:fig_8e}
		\end{subfigure}
	\end{minipage}
	\caption{a) Input signal to 2D FFT for $Ek = 0$; b) FFT output of $Ek = 0$ case; c) input signal to 2D FFT for $Ek = 339$; d) FFT output of $Ek = 0$ case; e) variation of axial and azimuthal wavenumbers, $\nu_z$ and $\nu_\theta$, with $Ek$. The rest of the parameters remain unaltered from Fig. 2.}
	\label{fig:fig8}
\end{figure}
The `angled' waves occur due to a competition between %
the axial flow due to gravity and the azimuthal flow due to the rotation of the cylindrical surface.
On this basis, we propose an expression that provides  an estimate of the angle, $\phi_a$, which assumes the
direction of mean wave motion to be aligned with that of the resultant of 
axial and azimuthal characteristic velocities, ${\cal{U}}{\bf i}_z$ and ${\cal V}{\bf i}_\theta$, respectively: 
\begin{equation}
\phi_a = %
\tan^{-1}\left(\frac{\cal{U}}{\cal{V}}\right),  
\label{eq:phi_a}
\end{equation}
where ${\bf i}_z$ and ${\bf i}_\theta$ denote the unit vectors in the axial and azimuthal directions, respectively, ${\cal V}=\Omega R$, and a reasonable estimate for $\cal{U}$ is given by the wave celerity that can be extracted readily from the time-space plots in the 2D simulations. 
In Fig. \ref{fig:fig_7f}, we plot the variation of $\phi$ with $Ek$, where it can be seen that $\phi$ decays as a function of $Ek$. %
If one were to choose the Nusselt solution, $u_N$, for ${\cal{U}}$ then using the definitions for $Re$ and $Ek$, it is seen readily that $\tan \phi_a = Re/Ek$, thus 
$\tan \phi_a \sim Ek^{-1}$ for fixed $Re$. 
This corresponds to the assumption that the %
falling film has %
a uniform film thickness. %
As can be seen from Fig. \ref{fig:fig_7f}, this approximation gives rise 
to a similar qualitative trend to that already discussed but a significant quantitative discrepancy; the latter is attributed %
to the presence of the waves, which are not taken into account in the Nusselt solution. Also shown in Fig. \ref{fig:fig_7f} is the prediction from Eq. (\ref{eq:phi_a}) utilising the wave celerity, $c$, whereby $Re_c = \frac{\rho c h_N}{\mu}$, which shows
excellent agreement with those obtained from the 3D numerical simulations.

We perform two-dimensional FFTs of the interface contour in the 
$(\theta,z)$ plane, as shown in Fig. \ref{fig:fig_8a}, with 
the contour plot of the power spectra shown in Fig. \ref{fig:fig_8b}; here,
$\nu_z$ and $\nu_\theta$ denote the wavenumbers in the axial and azimuthal directions, respectively.
The discrete Fourier transform $Y$ of an $m$-by-$n$ matrix $X$ is given by:
\begin{equation}
Y_{p+1, q+1}=\sum_{j=0}^{m-1} \sum_{k=0}^{n-1} \omega_{m}^{j p} \omega_{n}^{k q} X_{j+1, k+1},
\end{equation}
\noindent where $\omega_m$ and $\omega_n$ are complex roots of unity:
\begin{equation}
\omega_{m}=e^{-2 \pi i / m}, \quad \omega_{n}=e^{-2 \pi i / n},
\end{equation}
\noindent $i$ is the imaginary unit, $p$ and $j$ are indices that span 0 to $m-1$, and $q$ and $k$ are indices that span 0 to $n-1$.
Here, the $m\times n$ input is the matrix of film heights in the $(\theta, z)$ plane. We require similar modifications to the input as in the 2D case, in order to avoid biasing the result via the waveless section of the flow or through the non-zero mean amplitude bias inherent within the FFT. As seen in Fig. \ref{fig:fig_8a}, we modify the input using $L_{2D}$ and $\tilde{h} - \bar{\tilde{h}}$, for this case $Ek = 0$. After applying the FFT, the result is mapped onto the $(\nu_\theta,\nu_z)$ plane, as shown in Fig. \ref{fig:fig_8b}, where the peak of maximum amplitude is highlighted in yellow. From Fig. \ref{fig:fig_8a}, we can qualitatively estimate 8 waves within the domain yielding $\nu_z \approx 70 \ \text{m}^{-1}$. This coincides with $\nu_z$ readings computed from the 2D simulation and the FFT result in Fig \ref{fig:fig_8b}. %
We also note in this no-rotation, $Ek = 0$, case, that the waves are not angled, %
which is reflected in the FFT result as $\nu_\theta = 0$ at the primary peak, showing the absence of a contribution from the azimuthal wavenumber $\nu_\theta$. As $Ek$ is increased, the waves become angled leading to a non-zero contribution to $\nu_\theta$. An example of this is provided by the $Ek = 339$ case for which the interface contour is shown in Fig. \ref{fig:fig_8c}. %
The associated FFT output is shown in Fig \ref{fig:fig_8d}, with the peak at $\nu_z = 63 \ \text{m}^{-1}$ and $\nu_\theta = 16 \ \text{m}^{-1}$. %
As demonstrated above, the angles of the waves strongly depend on $Ek$ and we expect a similar relationship between $Ek$ and $\nu_\theta$. In a similar manner to the results of the FFT for the 2D simulation for which $\nu_z$ is weakly-dependent on $Ek$ (see Fig. \ref{fig:fig_5f}), Fig. \ref{fig:fig_8e} shows that while %
$\nu_z$ remains %
approximately constant with $Ek$, $\nu_\theta$ decreases for $Ek > 0$ and follows a similar trend to that shown in Fig. \ref{fig:fig_7e} for angle $\phi$ vs $Ek$. %
One expects the length scale of a coherent structure in the azimuthal direction, $\lambda_\theta$, to be the product of a linear velocity set by the cylinder rotation, $\Omega R$, and a time scale set by the gravitational forces, $h_N/u_N$. From the definitions of $Re$ and $Ek$, $\lambda_\theta \sim \Omega R h_N/u_N \sim (Ek/Re)h_N$. Thus, $\lambda_\theta \sim Ek$ and $\nu_\theta \sim \lambda_\theta^{-1} \sim Ek^{-1}$ for fixed $Re$.

\begin{figure}
	\begin{subfigure}{0.49\linewidth}
	\centering
	\caption{}
	\includegraphics[width = \linewidth]{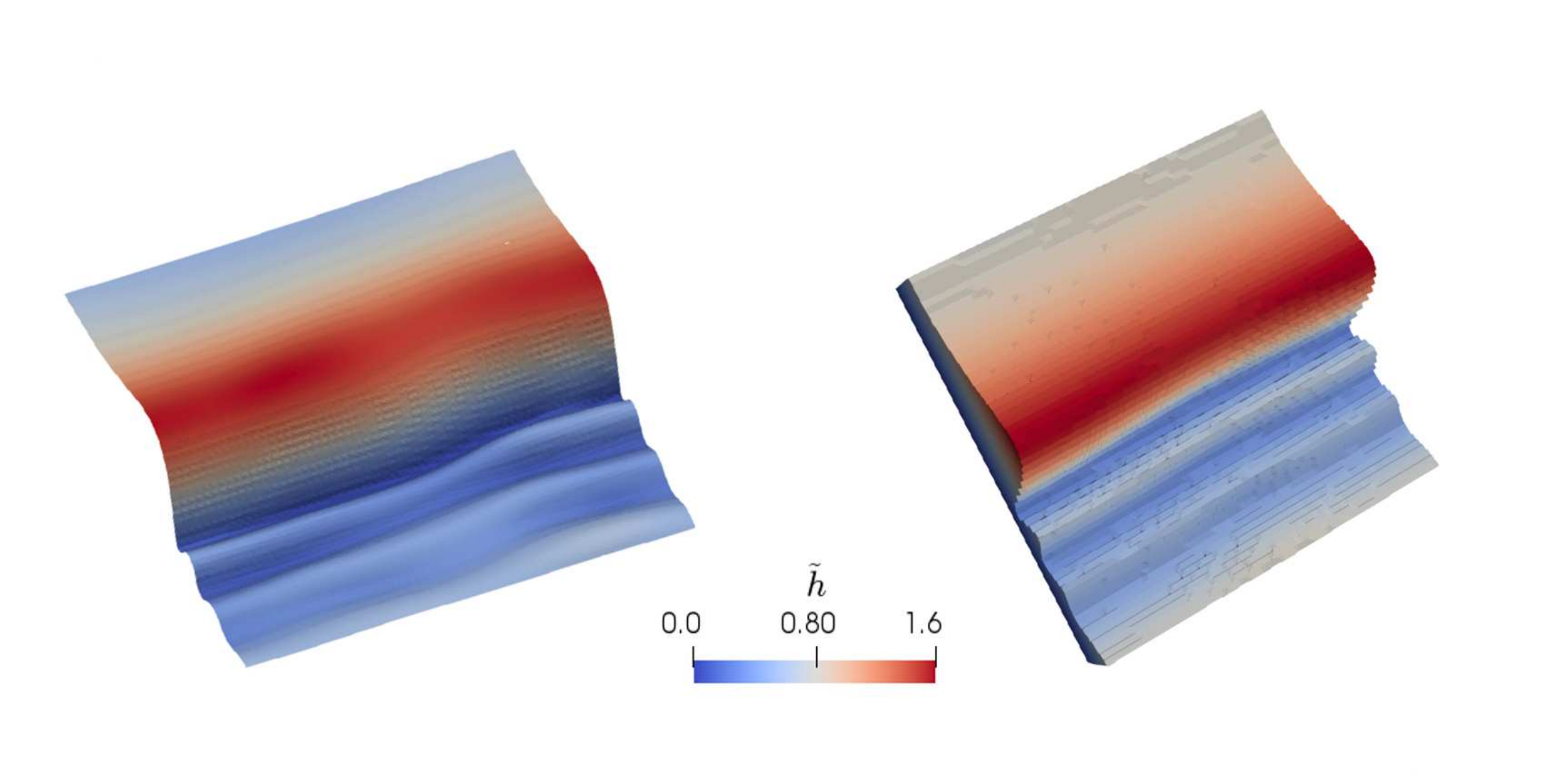}
	\label{fig:fig_10a}
	\end{subfigure}
	\begin{subfigure}{0.49\linewidth}
	\centering
	\caption{}
	\includegraphics[width = \linewidth]{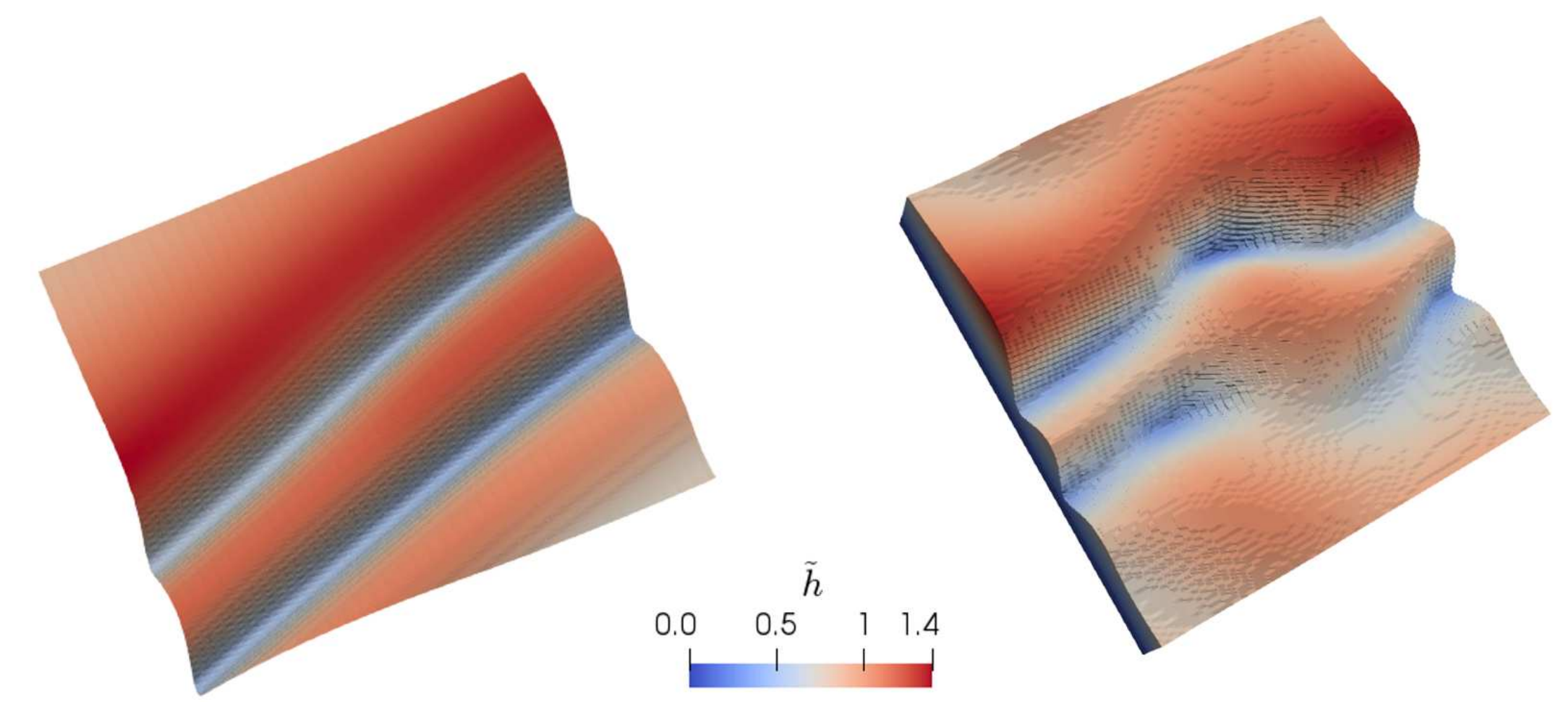}
	\label{fig:fig_10b}
	\end{subfigure}
	\caption{Comparison of 3D simulations for a) $Ek = 0$ and b) $Ek$ = 339. In each panel the VOF and WRIBL solutions are shown in the left and right panels, respectively. The rest of the parameters are unaltered from Fig. 2.}
	\label{fig:fig10}
\end{figure}

The full second-order WRIBL can be run in 3D space in order to draw a comparison with the VOF simulations, as in Fig. \ref{fig:fig6_1}. The 3D domain size was selected in order to obtain a single wave in the domain, similar to the 2D case, with the exception that it imposes double periodic boundary conditions. Hence, accounting for the wavenumber computed from the FFT, a 0.0143 m square domain was selected. The analogy with a falling film on an inclined plate proposed for the 2D analysis was also employed for this 3D configuration. As stated in section \ref{section_3} the comparison is valid for $1/\beta >> 10$ \cite{chen_ijhmt_2004} as is the case here.  A developed wave profile from the VOF simulation is compared to that obtained from the WRIBL model in Fig. \ref{fig:fig10}, using the same domain area. The variation in film height in the the $Ek = 0$ case is much more significant, reflecting the enhanced stability due to rotation. Furthermore, the capillary waves are smaller and more numerous, reflected in both simulations. From Fig. \ref{fig:fig_10b}, we observe the angled waves are not captured using the WRIBL model, as this uses a normal acceleration compared to a physically moving wall. Comparing the maximum film height to the 2D case in \ref{fig:fig6_1} we note that $h_{max}$ is systematically $5 \%$ lower in the 3D case, possibly due to spanwise surface tension interactions. Despite this, after applying the criteria of $h_{max}/h_N > 1.5$ for recirculation, we yield the same result, with sufficient $Ek$ suppressing wave formation and recirculation, reflected in Fig. \ref{fig:fig10}. While there is a clear difference in the wave orientation with the rotationally-resolved DNS in Fig. \ref{fig:fig_10b}, the ability of the WRIBL method to predict the effects of rotation on wave formation and recirculation in a developed state appears advantageous for rapidly establishing operating conditions for the intensification and control of convective transport within the film. 

\section{Conclusion}\label{section_5}
We have investigated the flow of a thin film falling under gravity on the inside of a vertical cylinder undergoing steady rotation. We have studied the interfacial dynamics in two and three dimensions via numerical simulations using the volume-of-fluid (VOF) method. We have shown that the cylinder rotation, characterised by an Ekman number $Ek$, has a significant effect on the structure of the interfacial waves.  Increasing the relative magnitude of the %
centrifugal force enhances interfacial stability, suppresses the wave formation brought about by the Kapitza instability, %
and extends the development length from the domain inlet prior to the appearance of the waves. %
Furthermore, our three-dimensional simulations demonstrate the formation of angled waves where the angles are  
a consequence of the resultant of the axial and azimuthal velocities. We have also compared the predictions from a Matlab-based weighted residual integral boundary layer method approach %
\cite{rohlfs_swx_2018} to the two- and three-dimensional VOF results and used it to construct a phase diagram, in which we highlighted the regions in Ekman-Reynolds number space wherein recirculation within the wave peaks is expected to arise.

\vspace{0.2in}

O. K. M. acknowledges funding from PETRONAS and the Royal Academy of Engineering for a Research Chair in Multiphase Fluid Dynamics, 
and from the Engineering and Physical Sciences Research Council UK through the MEMPHIS (EP/K003976/1) and PREMIERE (EP/T000414/1) Programme Grants. J. S. acknowledges funding from the European Union's Horizon 2020 research and innovation programme under the Marie Sk\l{}odowska-Curie Individual Fellowship grant agreement No. 707340.

\end{document}